\documentclass[11pt]{article}
\usepackage[margin=1in]{geometry}             
\usepackage{amssymb,amsmath,amsthm,amsfonts,mathrsfs}
\usepackage{graphicx, graphics, color}
\usepackage{tikz,mathpazo}
\usepackage{multirow}
\usepackage{color, colortbl}
\usepackage{caption,subcaption}
\usepackage[linesnumbered,ruled]{algorithm2e}
\usepackage{authblk}
\usepackage{epstopdf}
\usepackage{verbatim}
\usepackage{bm}
\usepackage[colorinlistoftodos]{todonotes}
\usetikzlibrary{shapes.geometric, arrows}
\usepackage{indentfirst}
\usepackage[first=0,last=9]{lcg}

\usepackage{xcolor}

\definecolor{Gray}{gray}{0.9}
\definecolor{LightCyan}{rgb}{0.88,1,1}
\usepackage[utf8]{inputenc}
\usepackage{longtable}
\usepackage{siunitx} 
\usepackage{natbib}
\title{Gaussian Process Assisted Active Learning of Physical Laws}
\author[1]{Jiuhai Chen}
\author[1]{Lulu Kang}
\author[2]{Guang Lin}
\affil[1]{Department of Applied Mathematics, Illinois Institute of Technology}
\affil[2]{Department of Mathematics, Department of Mechanical Engineering, Department of Statistics(Courtesy), Department of Earth, Atmospheric, and Planetary Sciences(Courtesy), Purdue University}

\begin{document}
\maketitle

\begin{abstract}
In many areas of science and engineering, discovering the governing differential equations from the noisy experimental data is an essential challenge.
It is also a critical step in understanding the physical phenomena and prediction of the future behaviors of the systems.
However, in many cases, it is expensive or time-consuming to collect experimental data.
This article provides an active learning approach to estimate the unknown differential equations accurately with reduced experimental data size.
We propose an adaptive design criterion combining the D-optimality and the maximin space-filling criterion.
In contrast to active learning for other regression models, the D-optimality here requires the unknown solution of the differential equations and derivatives of the solution.
We estimate the Gaussian process (GP) regression models from the available experimental data and use them as the surrogates of these unknown solution functions.
The derivatives of the estimated GP models are derived and used to substitute the derivatives of the solution.
Variable selection-based regression methods are used to learn the differential equations from the experimental data.
Through multiple case studies, we demonstrate the proposed approach outperforms the D-optimality and the maximin space-filling design alone in terms of model accuracy and data economy.
\\
\noindent{\bf Keywords:} Active learning; D-optimal design; Gaussian process model; Sequential design; Space-filling design; Variable selection.
\end{abstract}

\section{Introduction}

A wide variety of physical phenomena such as sound, heat, electrostatics, electrodynamics, fluid dynamics, elasticity, or quantum mechanics, are governed by physical laws that are often described by differential equations.
Thus, differential equations, such as ordinary differential equations (ODE) and partial differential equations (PDE), play an important role in many areas of science and engineering.
However, for many complex systems, it is difficult for researchers to deduce the governing equations from noisy data.
Therefore, discovering the governing equations from noisy data is an essential task in many sciences and engineering disciplines, and is critical to the understanding of physical phenomena and prediction of the future behaviors of the systems under study.

There have been many methods developed to achieve this goal.
Among them, one earlier approach was delivered by \cite{bongard2007automated}.
It was the first method that can automatically generate symbolic equations for a nonlinear coupled dynamical system directly from time-series data.
Pursuing the same direction, quite a few new ideas have been introduced.
\cite{brunton2016discovering} used sparse regression to determine the terms in the dynamic equations.
Following this idea, \cite{schaeffer2017learning} applied the shrinkage method and minimized the $L_1$-norm regularized least squares to identify the underlying PDE.
\cite{long2017pde} introduced a new feed-forward deep neural network, called PDE-Net, to accurately predict the dynamics of complex systems and to uncover the underlying hidden PDE models.
More recently, \cite{zhang2018robust} proposed to select candidate terms for the underlying equations using dimensional analysis and approximate the weights of the terms using threshold sparse Bayesian regression.

These works have significantly advanced the progress of data-driven modeling of differential equations.
But they are all based on a large quantity of data.
Especially for the PDE-net method, a huge amount of data is required to train the neural network.
One exception in the existing literature is introduced by \cite{raissi2018hidden}.
Their method does not require a large amount of data, as it leverages the underlying laws of physics, meaning that the time-dependent PDEs are assumed to be known.
The main task of learning is to identify a few unknown parameters in the known equations.
As effective as this approach is, it is not applicable when the explicit form of the time-dependent PDEs are unknown to the experimenter.

In many disciplines, data collection, or experimentation, takes time and resources.
When a researcher cannot afford the experiment's cost, the insufficient data could lead to incorrect mathematical models.
On the contrary, if the researcher collects more data than necessary, it would cause a waste of time and resources.
Without knowing how much data are required, either scenario is likely to occur.
This practical challenge and how sequential approaches can be used to overcome it is well-demonstrated in Section 5. 

In Section 5, we illustrate an air pollution monitoring application, where the data are collected via sensors.
For example, in \cite{cheng2011modeling}, each sensor measures the concentration of the pollutant, such as carbon monoxide, or CO.
Sensors, such as CO monitors, can be expensive.
If all the data are collected in a single trial, the experimenter requires a sufficient number of sensors to take measurements from different spatial locations spreading out across the domain.
Fortunately, the sensors in this scenario are mobile, as shown in \cite{cheng2011modeling}.
The experimenter can quickly move the sensors (by manpower or automation) to the new locations and collect a new batch of data.
The compromising assumption is that the data collected sequentially at very short time intervals can be approximately considered from the same time point.
It is reasonable to assume so as long as the diffusion process is slow enough and does not change significantly in a short time period.
But if this assumption does not stand, alternatively, the experimenter can restart the diffusion process and move the sensors to the new locations and collect a new batch of data at the same time point.

We propose an active learning approach that combines the optimal design method and 'the variable selection technique, to identify the significant terms in the mathematical equations.
The optimal design criterion combines the maximin space-filling criterion and the D-optimality.
The latter ensures the accurate estimation of the differential equations by linear regression.
However, the D-optimality involves the equations' unknown solution functions and their unknown derivatives, and thus we substitute them via the Gaussian process (GP) surrogate models and their derivatives.
This is why we also need the design to be space-filling so that it can explore the design space more thoroughly to fit the GP models.
The weights of combining the two criteria are calculated adaptively from the currently estimated differential equations and the GP models.
We give the adaptively combined D-optimal and space-filling criterion an acronym ACDS.
Details are explained in Section 3.
Through case studies in Section 4 and 5, we show that the proposed method outperforms the space-filling and D-optimal design alone in terms of model accuracy and economy of the experimental run size.
More remarks on the case studies are elaborated in Section 6.
The paper is concluded with some discussion in Section 7.
The codes and data are available from \url{https://github.com/ACDS-code/ACDS.git}.

\section{Discovery of Physical Law}

\subsection{General Review}

There are many physical laws represented by various kinds of differential equations.
In the scope of this paper, we only consider differential equations of the form in PDEs \eqref{eq:PDEs} and ODEs \eqref{eq:ODEs}.
Specially, the type of PDEs we focus on is
\begin{equation}
\label{eq:PDEs}
\frac{\partial \bm u}{\partial t}=\bm f(\bm x, \mathscr{L}_{\bm x}\bm u), \quad \bm x\in\Omega, t\in[0,T],
\end{equation}
where $\bm u(\bm x, t)\in \mathbb{R}^d$ denotes the state of a system at time $t$, i.e., the solution of \eqref{eq:PDEs}, $\bm x\in \mathbb{R}^p$ represents other variables required to specify the state of the system, such as the spatial location in the system, $\Omega\subset \mathbb{R}^p$ and $[0, T]$ are the domain of $\bm x$ and time in which the equations are established, and $\mathscr{L}_{\bm x}$ is a linear or nonlinear operator applied to $\bm u$.
The subscript in $\mathscr{L}_{\bm x}$ denotes that the differentiation is in $\bm x$.
The function $\bm f$ is a vector of \emph{polynomial functions} in $\mathbb{R}^d$ and has the input $\bm x$ and $\mathscr{L}_{\bm x} \bm u$.
The operator $\mathscr{L}_{\bm x}$ and the function $\bm f$ together define the dynamic constraints of the systems.
The explicit form of $\bm f$ and $\mathscr{L}_{\bm x}$ are unknown and are the target of learning from experimental data.
Following the PDE learning approach proposed by \cite{raissi2018hidden}, we restrict that $\bm f(\bm x, \mathscr{L}_{\bm x}\bm u)$ does not contain any polynomial terms of $t$ variable.
With this assumption, we only need data at a particular time point, $t=t_s$, to learn $\bm f(\bm x, \mathscr{L}_{\bm x}\bm u)$. 

The system of ODEs can also be expressed by a simpler version of \eqref{eq:PDEs}.
The state of the system $\bm u(t)$ only depends on the variable $t$, and the system of ODEs is
\begin{equation}
\label{eq:ODEs}
\frac{d \bm u}{d t} = \bm f(t, \bm u), \quad t\in [0, T],
\end{equation}
where $\bm f$ is the governing function of the system dynamics.
We assume $\bm f$ is a vector of \emph{polynomial functions} of $t$ and $\bm u$.
For ODEs, since $t$ is the only input variable of $\bm u$, it does not matter if $\bm f(t, \bm u)$ explicitly contains any terms of $t$.

A wide range of physical laws can be represented by the types of PDEs \eqref{eq:PDEs} and ODEs \eqref{eq:ODEs}. 
One such PDE example of \eqref{eq:PDEs} is the classic heat equation 
\[
\frac{\partial u}{\partial t}=\alpha\left(\frac{\partial^2 u}{\partial x_1^2}+\frac{\partial^2 u}{\partial x_2^2}+\frac{\partial^2 u}{\partial x_3^2}\right).
\]
It describes how the distribution of some quantity, such as heat, evolves over time in a homogeneous and isotropic medium. 
The function $u(\bm x, t)$ is the temperature of location $\bm x$ at time $t$.
Another ODE example of \eqref{eq:ODEs} is the kinematic equation, which models the free-falling object problem. 
Assume $L$ is the displacement, $g$ stands for the acceleration of the object, and $v_0$ is the initial velocity.
The kientmatic equation is $\frac{d L}{d t} = v_0 + g\times t$. 
We also show some other famous PDE and ODE examples in Section 4 and 5. 

To explain the general data-driven modeling framework, we use a simple PDE as an example.
\begin{equation*}
\frac{\partial u}{\partial t}=f(u,u_x), x\in[a,b], t\in[0,T].
\end{equation*}
Both $u$ and $f$ are one-dimensional functions.
At time $t=t_s$, the observed data are
\[\left\{x_i,u_i,(\frac{\partial u}{\partial t})_i,(\frac{\partial u}{\partial x})_i\right\}_{i=1}^N,\]
where $u_i=u(x_i, t_s)$, $(\frac{\partial u}{\partial t})_i= \frac{\partial u}{\partial t}(x_i, t_s)$, $(\frac{\partial u}{\partial x})_i= \frac{\partial u}{\partial x}(x_i, t_s)$.
As pointed out by \cite{raissi2018hidden}, we do not need the observations at other time points to estimate $f(u, u_x)$ because it does not involve the variable $t$, which greatly reduces the amount of data required.
In the framework introduced by \cite{bongard2007automated} and many other following ones, $f(u, u_x)$ is assumed to be a linear combination of some terms (or bases).
Linear regression combined with variable selection methods is used to identify the significant terms from a group of preset candidate terms.
The linear coefficients of these terms are estimated in the process.
For this example, we pick the set of candidate basis functions to be $\{1,u,(\frac{\partial u}{\partial x}),u^2,(\frac{\partial u}{\partial x})^2,u(\frac{\partial u}{\partial x})\}$.
The linear regression is applied to the following model.
\begin{equation}\label{eq:reg_setup}
      \begin{bmatrix}
      (\frac{\partial u}{\partial t})_1  \\
      (\frac{\partial u}{\partial t})_2  \\
      \vdots
\\
      (\frac{\partial u}{\partial t})_N
      \end{bmatrix}=
      \begin{bmatrix}
      1 & u_1 & (\frac{\partial u}{\partial x})_1 & u_1^2 & (\frac{\partial u}{\partial x})_1^2 & u_1(\frac{\partial u}{\partial x})_1  \\
      1 & u_2 & (\frac{\partial u}{\partial x})_2 & u_2^2 & (\frac{\partial u}{\partial x})_2^2 & u_2(\frac{\partial u}{\partial x})_2  \\
      \vdots & \vdots & \vdots & \vdots & \vdots & \vdots   \\
      1 & u_N & (\frac{\partial u}{\partial x})_N & u_N^2 & (\frac{\partial u}{\partial x})_N^2 & u_N(\frac{\partial u}{\partial x})_N  \\
      \end{bmatrix}
      \begin{bmatrix}
        \beta_0
\\
        \beta_1
\\
        \beta_2   \\
        \beta_3   \\
        \beta_4    \\
        \beta_5    \\
        \end{bmatrix}+
        \begin{bmatrix}
          \epsilon_1  \\
          \epsilon_2  \\
          \vdots    \\
          \epsilon_N   \\
          \end{bmatrix},
\end{equation}
where $\bm \epsilon=[\epsilon_1,\epsilon_2,\cdots,\epsilon_N]^T$ is the model error.
Different factors can contribute to creating the model error, such as model inadequacy and measurement noise.
Numerical errors are also likely to occur when some of the derivatives are not observed but calculated by finite-difference from observations.
It is difficult to quantify how the errors and noise contained by $\{u_i,(\frac{\partial u}{\partial t})_i,(\frac{\partial u}{\partial x})_i\}_{i=1}^N$ are aggregated in $f(u, u_x)$.
Therefore, for simplicity, all the existing methods assume $\bm \epsilon \sim N(0,\sigma^2\bm I_N)$.
The data-driven modeling is to estimate $\bm \beta=[\beta_0,\beta_1,\beta_2, \beta_3,\beta_4,\beta_5,\beta_6]^\top$ with certain sparsity.

To sum up, the proposed active learning methods can be used to recover the underlying differential equations taking the form of \eqref{eq:PDEs} and \eqref{eq:ODEs}.
They satisfy (1) $\bm f$ is a vector of polynomial functions of their inputs; (2) for PDEs, $\bm f$ cannot explicitly contain any terms involving $t$, and the differential operator $\mathcal{L}_{\bm x}$ is only applied to $\bm x$; (3) for ODEs, $\bm f$ does not contain any derivative terms of $\bm u$.

\subsection{Candidate set of basis functions}

In general, the preset candidate of basis functions should be large enough to include the actual terms contained by the underlying differential equations.
Domain knowledge is certainly helpful to construct the basis functions.
In \cite{zhang2018robust}, the authors illustrated using tensor product to construct the basis functions as follows
\begin{equation}\label{eq:candidates}
  \bigotimes^{k_1}\left\{1, \bm x, \bm u, \frac{\partial u_1}{\partial x_1},\ldots, \frac{\partial u_d}{\partial x_1}, \ldots, \frac{\partial u_1}{\partial x_p},\ldots, \frac{\partial u_d}{\partial x_p}, \ldots, \right\},
\end{equation}
where the second ellipsis represents the partial derivatives of $u_i$ for $i=1, \ldots, d$ to certain elements of $\bm x$ up to a user specified order $k_2$.
The operation $\bigotimes^{k_1} S$ denote tensor product of $k_1$ copies of set $S$.
For example, assume $p=2$, $k_1=1$, and $k_2=2$, and then the candidate set is
\begin{align*}
&\left\{1, x_1, x_2, u, \frac{\partial u}{\partial x_1},\frac{\partial u}{\partial x_2},
\frac{\partial^2 u}{\partial^2 x_1}, \frac{\partial^2 u}{\partial^2 x_2}, \frac{\partial^2 u}{\partial x_1\partial x_2}\right\}.
\end{align*}
In another example, let $p=1$ and $k_1=k_2=2$, and the candidate set is
\begin{align*}
\bigotimes^2 &\left\{1, x, u, \frac{\partial u}{\partial x},\frac{\partial^2 u}{\partial^2 x}\right\}.\\
=&\left\{1, x, u, \frac{\partial u}{\partial x},\frac{\partial^2 u}{\partial^2 x}, x^2, xu, x\frac{\partial u}{\partial x},x\frac{\partial^2 u}{\partial^2 x}, u^2, u\frac{\partial u}{\partial x}, u\frac{\partial^2 u}{\partial^2 x}, \left(\frac{\partial u}{\partial x}\right)^2,\frac{\partial u}{\partial x}\frac{\partial^2 u}{\partial^2 x},
\left(\frac{\partial^2 u}{\partial^2 x}\right)^2\right\}.
\end{align*}
Clearly, the tensor product can easily construct a large pool of basis functions.
\cite{zhang2018robust} then proposed to screen the basis functions by comparing the ``dimensionality'' of the two sides of the equation.
For instance, if the unit of $\frac{\partial u}{\partial t}$ is meter per second, then the units of all the basis functions should also be meter per second.
Any terms having different units (or dimensions in physics) should be screened out from the pool of candidates.

\subsection{Variable selection}

As reviewed in Section 1, various methods have been proposed to estimate the linear coefficients, $\bm \beta$.
Essentially, it is a problem of variable selection for the linear regression model.
Many existing methods can be used together with the later proposed active learning approach.
We have tried three variable selection methods.
They are the best subset selection \citep{Beale1967The,Hocking1967Selection}, stepwise selection \citep{Draper1966Applied}, and shrinkage methods like Lasso \citep{Tibshirani1996Regression}.
The best subset selection we have tried is the formulation of \cite{Bertsimas2015BestSS}, which turns the variable selection into a mixed-integer programming problem.
Based on our investigation, we choose the forward stepwise regression combined with the Bayesian information criterion (BIC) as the variable selection method to illustrate the proposed active learning approach.
Here are the reasons.

First, forward stepwise regression is easier to implement and faster to compute than the best subset selection by \cite{Bertsimas2015BestSS}, even though the two have similar performances.
Second, BIC returns sparser regression models than some other criteria such as AIC, and it suits the purpose of learning differential equations since most underlying differential equations have few terms.
More importantly, as shown in our comparison with Lasso in Figure \ref{Lasso}, the forward stepwise regression combined with BIC is more accurate than Lasso in terms of identifying correct terms.
This point is also illustrated in \cite{zhang2018robust}, in which the authors proposed a new variable selection that outperforms Lasso.
We admit that the stepwise regression might not perform well in the face of strong collinearity, and it could introduce biases since it is a greedy search.
But these issues have not shown up in our studies.
Ultimately, the specification of a variable selection method is not the primary focus of this paper, and we encourage readers to choose the suitable one for their applications.

\section{Active Learning}

\subsection{Motivation of a new design criterion}

The active learning is also known as the sequential experimental design method in statistics.
Various versions and different applications of active learning have been introduced.
The early works include \cite{chernoff1959sequential} and \cite{blot1973sequential}.
Recent ones can be found in \cite{williams2000sequential,lin2004sequential,dror2008sequential,dasgupta2008statistical,deng2009active}, etc.
In general, active learning consists of the following steps.
\begin{enumerate}
\item [Step 1] Construct an initial design, such as space-filling design, collect the data, and build an initial model.
\item [Step 2] Based on the current fitted model, update the user-specified design criterion, and select the next batch of design points by optimizing the criterion.
\item [Step 3] Collect the data and update the model.
\item [Step 4] Iterate Steps 2 and 3 until the stop condition is satisfied.
\end{enumerate}

The design criterion used in Step 2 should fit the purpose of the experiment.
In our case, the accuracy of the estimated coefficients of the linear regression model is crucial.
A model-based optimal design criterion can be used \citep{doi:10.1002/wics.100}.
Among the various optimal designs, the D- and A- optimal design focuses on the variance of the estimated coefficients.
We choose the more widely used D-optimal design to select the design points in variable $\bm x$.

For a regular linear regression model, the D-optimal design maximizes $\det(\bf {M}^\top \bf {M})$ with respect to the design points, where $\bf {M}$ is the $N\times k$ model matrix of $k$ basis functions evaluated at the $N$ design points.
The $k$ basis functions are the model terms specified by the experimenter.
Their values at the potential design points can be easily calculated.
But this is not the case for learning differential equations, where the candidate basis functions involve the unknown solution of the differential equations and its derivatives.
For instance, the basis functions in the example in \eqref{eq:reg_setup} include $\{u, u_x, u^2, (u_x)^2, uu_x\}$.
In the process of active learning, we only have observations of $u$, $u_x$ and $u_t$ at the existing design points (at time $t=t_s$), but not at the potential design points.

To construct the model matrix $\bf {M}$, we need to evaluate $\bm u(\bm x,t)$ and its derivatives at the potential design points at time $t=t_s$.
One option is to solve the currently estimated version of the differential equations.
But this can be prohibitively difficult because the estimated differential equations still contain a large number of terms when only a few data are collected.
Some terms, such as the higher-order derivatives of $\bm u(\bm x,t)$ or the products between derivatives, might not be contained by the true differential equations, but are not yet screened out in the early iterations.
They make the differential equations complex and computational to solve.
Moreover, the early estimated differential equations are more likely to differ from the true equations significantly.
As a result, the solution $\bm u(\bm x, t)$ would behave differently from the true system in the unexplored design space.
The derivatives of the solution might diverge further from the true derivatives.
Therefore, even if we can solve the estimated differential equations in the early stages of active learning, the solution could lead to the ``wrong'' design points for the subsequent learning.

Alternatively, we can build a surrogate model of $\bm u(\bm x,t_s)$ based on the current available observations $\{\bm x_i, \bm u(\bm x_i,t_s)\}_{i=1}^{n}$ for $i=1,2,\ldots,n$, where $n$ is the currently available sample size.
The surrogate model is an empirical statistical model that is often used to analyze the outputs from computer experiments or simulations, in which the functional relationship between the input variables and outputs is complex and highly nonlinear.
For example, many computer experiments are run through complex numerical PDE solvers.
Among all statistical modeling methods, Gaussian Process (GP) regression, also known as \emph{kriging}, has been widely used for computer experiments \citep{santner2003design} for several reasons.
First, due to the mathematical simplicity of the GP assumption, it is relatively easy to obtain the prediction and statistical inference.
Second, the GP predictor with nugget effect (or the posterior mean if Bayesian framework is used) is identical to the kernel ridge regression based on reproducing kernel Hilbert space (RKHS) \citep{kanagawa2018gaussian}.
Therefore, the GP regression possesses the same theoretical properties of RKHS regression which provides a clear analysis of the approximation error \citep{wendland2004scattered}.

We choose the GP regression as the surrogate model for $\bm u(\bm x,t_s)$ to construct the basis functions.
Besides the above reasons, we have a more important motive.
The properties of the covariance function around $\bm x=0$ determine the smoothness properties of the GP.
So we only need to choose the proper kernel as the covariance function to match the smoothness of GP to $\bm u(\bm x,t)$.
Thanks to this property, we can first build the GP regression to replace $\bm u(\bm x,t=t_s)$ and then obtain the derivatives of the fitted GP model analytically, which are used to replace derivatives of $\bm u(\bm x,t=t_s)$ of $\bm x$.
Other statistical models, such as splines, are mostly based on low-order polynomial functions of $\bm x$.
If these methods are used, we need to build separate models for each of $\bm u(\bm x,t)$ and its derivatives, because the polynomials may not match the smoothness of the $\bm u(\bm x,t)$.

In the remaining section, we first introduce the ACSD design criterion, then review the GP model and derive its derivatives, and lastly elaborate the entire active learning procedure to identify the unknown differential equations.

\subsection{ACDS design criterion}

The classic D-optimal design criterion is $\det(\bf {M}^\top \bf {M})$.
It does not depend on the response observations.
If there have been $n$ design points in the design, the model matrix $\mathbf {M}_{n}$ contains $n$ rows and $k$ columns.
To add the next design point, the D-optimal design is the solution of the following maximization problem.
\begin{align*}
\bm x_{n+1}&=\textrm{arg}\max_{\bm x\in \Omega}\det(\mathbf {M}^\top_{n+1} \mathbf {M}_{n+1})=\textrm{arg}\max_{\bm x\in \Omega}\det(\mathbf {M}^\top_{n} \mathbf {M}_{n}+\bm m(\bm x) \bm m(\bm x)^\top)\\
&=\textrm{arg}\max_{\bm x\in \Omega} (1+\bm m(\bm x)^\top (\mathbf {M}^\top_{n} \mathbf {M}_{n})^{-1}\bm m(\bm x))\det(\mathbf {M}^\top_{n} \mathbf {M}_{n}),
\end{align*}
where $\bm m(\bm x)$ is the $k\times 1$ vector of basis functions evaluated at $\bm x$.
Since the previous $n$ design points have been chosen already, $\det(\mathbf {M}^\top_{n} \mathbf {M}_{n})$ is invariant with respect to $\bm x_{n+1}$, and thus shall be omitted from the objective function.
We need to find $\bm x_{n+1}$ such that
\[
\bm x_{n+1}= \textrm{arg}\max_{\bm x\in \Omega} (1+\bm m(\bm x)^\top (\mathbf {M}^\top_{n} \mathbf {M}_{n})^{-1}\bm m(\bm x)).
\]
When $n$ is small (but still larger than the number of columns), we can add a regularization term to mitigate the ill-conditioning problem.
\begin{equation}
\bm x_{n+1}= \textrm{arg}\max_{\bm x\in \Omega} ( 1+\bm m(\bm x)^\top (\mathbf {M}^\top_{n}\mathbf {M}_{n}+\rho \mathbf {I}_k)^{-1}\bm m(\bm x)).
\label{D-optimal}
\end{equation}
Here $\rho$ is the noise-to-signal ratio.
If $u(\bm x, t)$ is one-dimensional, roughly, $\rho$ can be computed by $\hat{\sigma}^2/ s^2$, where $\hat{\sigma}^2$ is the estimated variance of the linear regression model with current $n$ observations, and $s^2$ is the sample variance of the column of $\frac{\partial u}{\partial t}$.
If $\bm u(\bm x,t)$ is multi-dimensional, it is the average of the noise-to-signal ratio for each dimension of $\bm u(\bm x, t)$.

During the active learning process, the model matrix $\mathbf {M}_n$ can be updated by removing some insignificant columns of bases, as long as variable selection is performed whenever new data are collected.
But sometimes the variable selection is not reliable when only a small amount of data has been collected.
Certain columns that are contained by the true differential equations might be dropped by mistake, which misleads the subsequent data collection.
To avoid this possibility, we decide not to update the model matrix by removing any candidate columns from ${\bf M}_n$ throughout the active learning procedure.

As explained earlier, we need to build a GP regression model as the surrogate of $\bm u(\bm x,t)$ to construct the basis functions at the potential design points.
But D-optimal design alone cannot facilitate a reasonable estimation of the GP model, as the optimal design points are usually clustered at a few local regions in the whole design space.
It could lead to numerical issues and cause the covariance matrix of the GP to be ill-conditioned.
Besides, the fitted GP model will not be a globally-accurate surrogate if only a few regions are explored.

Space-filling design \citep{joseph2016space} has been used frequently in combination with the GP model for computer experiments.
The design points are spread through the entire design space measured by various design criteria.
Sequential design approaches, such as \cite{harari2014optimal} and \cite{binois2019replication}, iteratively update the GP model using newly collected data and then select the next design point(s) to optimize some criterion, such as the mean square prediction error of the GP prediction.

In general, the mean squared error (MSE) of the GP prediction is smaller if the design has better space-filling property.
\cite{loeppky2010batch} found out that the maximin-distance designs perform comparably well with sequential designs that aim to reduce the mean squared error of the GP model.
Therefore, we choose the maximin-distance criterion to measure the quality of the space-filling design.
To add the design points sequentially, maximin design selects the next design $\bm x_{n+1}$ that maximizes the minimum distance between $\bm x_{n+1}$ and the current design points \citep{JOHNSON1990131},
\begin{equation}
\bm x_{n+1}=\textrm{arg}\max\limits_{\bm x\in \Omega}\min_{i=1,\ldots,n}\textrm{dist}(\bm x,\bm x_i),
\label{space-filling}
\end{equation}
where $\textrm{dist}(\bm x,\bm x_i)$ is the chosen distance metric for $\Omega$.
We simply use Euclidean distance $||\bm x-\bm x_i||_2$.

The proposed sequential design criterion must consider two fronts, the linear regression part that learns the significant terms in the differential equations and the GP surrogate model part that construct all the basis functions at the potential design points.
So we combine the D-optimal criterion and the maximin-distance criterion into one by linear combination.
\begin{equation}
\bm x_{n+1}=\textrm{arg}\max_{\bm x\in\Omega}\left\{\alpha_1\left[\min_{i=1,\ldots,n}||\bm x-\bm x_i||_2^2\right]+\alpha_2\left[1+\bm m(\bm x)^\top (\mathbf {M}^\top_{n} \mathbf {M}_{n})^{-1}\bm m(\bm x)\right]\right\}.
\label{criterion}
\end{equation}
Here $\alpha_1$ and $\alpha_2$ are the weights, and will be specified later.
But to properly choose the weights, we need to scale the two criteria into the same range.
The tight upper bound for the D-optimality is
\begin{align*}
&U_D=\max_{\bm m\in\mathcal{F}} (1+ \bm m^\top(\mathbf {M}^\top_n \mathbf {M}_n)^{-1}\bm m),
\end{align*}
where $\mathcal{F}$ is the feasible region for all $\bm m(\bm x)$ and $\bm x\in \Omega$.
Rigorously, $U_D$ can be calculated via quadratic programming if $\mathcal{F}$ can be decided based on the surrogate model of $\bm u(\bm x,t)$.
Because $\min_{i=1,\ldots,n}||\bm x-\bm x_i||_2^2\leq \frac{1}{n}\sum_{i=1}^n ||\bm x-\bm x_i||_2^2$, an upper bound for the minimum distance is
\[
U_S=\max_{\bm x\in \Omega} \frac{1}{n}\sum_{i=1}^n ||\bm x-\bm x_i||_2^2,
\]
which can also be solved by quadratic programming.
To simplify the computation, we obtain $U_D$ and $U_S$ from the pool of potential design points, which can be seen as a heuristic optimal solution.
Including the upper bounds, the proposed design criteria is
\begin{equation}\label{eq:acds}
\bm x_{n+1}=\textrm{arg}\max_{x\in\Omega}\left\{\alpha_1 \left[\frac{\min_{i=1,\ldots,n}||\bm x-\bm x_i||_2^2}{U_S}\right]+\alpha_2\left[\frac{1+\bm m(\bm x)^\top (\mathbf {M}^\top_{n} \mathbf {M}_{n})^{-1}\bm m(\bm x)}{U_D}\right]\right\}.
\end{equation}

Intuitively, the weights $\alpha_1$ and $\alpha_2$ should adjust the balance between the two design criteria.
Ideally, such adjustment should be data-driven and thus we compute $\alpha_1$ and $\alpha_2$ as follows.
\begin{equation}\label{eq:alphas}
\alpha_1=\frac{\hat{\tau}_{cv}^2}{\hat{\tau}_{cv}^2+\hat{\sigma}^2},\quad \alpha_2=\frac{\hat{\sigma}^2}{\hat{\tau}_{cv}^2+\hat{\sigma}^2}.
\end{equation}
Here $\hat{\sigma}^2$ is the estimated variance for $\epsilon$ from the stepwise linear regression, and $\hat{\tau}_{cv}^2$ is the leave-one-out cross-validation error from GP model \citep{dubrule1983cross}, which can be calculated via
\[
y_i-\hat{y}_{\theta,i,-i}=\frac{(\tilde{\bm K}_{xx}^{-1}y)_i}{(\tilde{\bm K}_{xx}^{-1})_i},\quad
\hat{\tau}_{cv}^2=\sum_{i=1}^{n}\frac{(y_i-\hat{y}_{\theta,i,-i})^2}{n}.
\]
Note that when $\bm u$ is multi-dimensional, $\hat{\tau}_{cv}^2$ is the average of the leave-one-out cross-validation error from each GP model fitting each dimension of $\bm u$.

The weights defined in \eqref{eq:alphas} are automatically updated based on the goodness of fit of the GP model and the regression model in each iteration.
If $\hat{\sigma}^2$ is significantly larger than $\hat{\tau}^2_{cv}$, it indicates that among the two fitted model, it is more urgent to collect the subsequent observations to improve the linear regression fit.
Thus, $\alpha_2$ is significantly larger than $\alpha_1$, which makes the D-optimality dominate the combined criterion \eqref{eq:acds}.
Conversely, the space-filling criterion dominates the criterion \eqref{eq:acds} if $\hat{\tau}^2_{cv}$ is significantly larger than $\hat{\sigma}^2$.
We name the proposed criterion \eqref{eq:acds} and the weights \eqref{eq:alphas} \emph{adaptively combined D-optimal and space-filling} criterion, or \emph{ACDS} for short.

\subsection{Gaussian Process regression and its derivatives}

In this part, we review the GP regression model and derive its first and second-order derivatives.
Using the general notation, we observe the data $\{\bm x_i, y_i\}_{i=1}^n$ with $\bm x\in \Omega\subset \mathbb{R}^p$, and $y_i\in \mathbb{R}$ is the univariate response observation for $\bm x=\bm x_i$.
The GP assumption says
\[
y(\bm x)=\mu(\bm x)+Z(\bm x)+\epsilon, \quad \textrm{where } Z(\bm x)\sim GP(0, k(\cdot,\cdot)) \textrm{ and }\epsilon \sim N(0,\sigma_0^2).
\]
with
\begin{equation*}
k(\bm x_i,\bm x_j)=\tau^2\exp\left\{-\sum_{s=1}^p \frac{(x_{i,s}-x_{j,s})^2}{2\omega_s}\right\}.
\end{equation*}
For simplicity, we assume $\mu(\bm x)$ is an unknown constant $\mu$.
The bandwidth parameter $\omega_s$ is positive for $s=1,\ldots,p$.
The parameters $\bm \theta=(\mu,\tau^2,\bm \omega,\sigma_0^2)$ are estimated by maximizing the log-likelihood \eqref{likelihood} based on the data.
\begin{equation}
2\log L=-(\bm y-\mu {\bf 1}_n)^\top\tilde{\mathbf K}_{xx}^{-1}(\bm y-\mu {\bf 1}_n)-\log\det(\tilde{\mathbf K}_{xx})-\text{constant},
\label{likelihood}
\end{equation}
where $\bm y$ is the vector of the observations $y_i$'s, $\tilde{\mathbf K}_{xx}=\mathbf K_{xx}+\sigma_0^2\mathbf I$, and $\mathbf K_{xx}$ denotes the covariance matrix with entries $[\mathbf K_{xx}]_{i,j}=k(\bm x_i,\bm x_j)$.
Once the parameters are replaced by the maximum likelihood estimates, the conditional mean of the response $\hat{y}(\bm x^*)$ corresponding to new inquiry point $\bm x^*$ is given by,
\begin{equation}
\hat{y}(x^*)=\hat{\mu}+{\bm k}_{x^{*}x}^\top \tilde{\mathbf{K}}_{xx}^{-1}(\bm y-\hat{\mu}{\bf 1}_n).
\end{equation}
It gives the predictor formula of the GP surrogate model.
The vector $\bm k_{x^{*}x}^\top =[k(\bm x^*, \bm x_1),\ldots, k(\bm x^*, \bm x_n)]$ is the vector of covariance between $\bm x^*$ and $\bm x_i$'s, and $\hat{\mu}={\bf 1}_n^\top \tilde{\mathbf K}_{xx}^{-1}\bm y/{\bf 1}_n^\top \tilde{\mathbf K}_{xx}{\bf 1}_n$.
We omit to review the conditional variance of the GP predictor, as we do not need inference information of the GP predictor in the proposed active learning approach.
In the predictor $\hat{y}(x^*)$, only the vector $\mathbf{K}_{x^{*}x}$ contains the variable $\bm x^*$.
The first order derivatives of the surrogate model are
\begin{equation*}
\frac{\partial \hat{y}(\bm x^*)}{\partial x^*_j}=\sum_{i=1}^n\frac{\partial k(\bm x^*,\bm x_i)}{\partial x^*_j}[ \tilde{\mathbf{K}}_{xx}^{-1}(\bm y-\hat{\mu}{\bf 1}_n)]_{i}\quad\textrm{for }j=1,\ldots, p,
\end{equation*}
and
\begin{equation*}
\frac{\partial k(\bm x^*, \bm x_i)}{\partial x^*_j}
=-\frac{(x^*_j-x_{i,j})}{\omega_j}k(\bm x^*, \bm x_i).
\end{equation*}
The second order derivatives are
\begin{equation*}
\frac{\partial^2 \hat{y}(\bm x^*)}{\partial x^*_l \partial x^*_j}=\sum_{i=1}^n\frac{\partial^2 k(\bm x^*,\bm x_i)}{\partial x^*_l \partial x^*_j}[\tilde{\mathbf{K}}_{xx}^{-1}(\bm y-\hat{\mu}{\bf 1}_n)]_i\quad\textrm{for }j,l=1,\ldots, p,
\end{equation*}
where
\begin{equation*}
\frac{\partial^2 k(\bm x^*,\bm x_i)}{\partial x^*_{l}\partial x^*_{j}}
=\left( \frac{(x^*_j-x_{i,j})}{\omega_j}\frac{(x^*_l-x_{i,l})}{\omega_l}-\frac{1}{\omega_j}\delta_{lj}\right)k(\bm x^*,\bm x_i)
\end{equation*}
with $\delta_{lj}=1$ if $l=j$ and 0 otherwise.
The derivatives of the GP model with more general form can be found in \cite{NIPS2018_7919}.
Higher-order derivatives can be obtained similarly.
Please note that this review is for the univariate response case.
When $\bm u\in \mathbb{R}^d$ is multi-dimensional ($d>1$), we only fit the GP model to each dimension of $\bm u$ individually instead of fitting all the dimensions into one joint model.
Based on the application, readers can use the joint GP model for multivariate responses, such as Co-Kriging \citep{myers1982matrix}, but more parameters have to be estimated due to the correlation between responses.

\subsection{Active learning procedure}

We summarize the proposed active learning procedure into Algorithm \ref{alg:actlearn}.
In the for-loop of the algorithm, when a new design point is added, a short cut formula
\[
(\bm A+\bm a\bm a^\top)^{-1}=\bm A^{-1}-\frac{\bm A^{-1}\bm a \bm a^\top \bm A^{-1}}{1+\bm a^\top \bm A^{-1}\bm a}
\]
can be used to update $(\mathbf {M}_{n+j-1}^\top \mathbf {M}_{n+j-1})^{-1}$ to $(\mathbf {M}_{n+j}^\top \mathbf {M}_{n+j})^{-1}$.
Also in the for-loop, the new rows in $\mathbf {M}_{n+j}$ are the basis functions of the selected design points, which are calculated based on the GP surrogate model and its derivatives.
Once the new data are collected, the newly added rows of the model matrix $\mathbf {M}_n$ need to be updated using the actual observations.

\begin{algorithm}
\SetKwInOut{Input}{Input}
\SetKwInOut{Output}{Output}
\Input{Prescribed tolerance of convergence $Tol$, the maximum allowed sample size $N_{\max}$, and the batch size $B$.}
\Output{Estimated PDE or ODE system}
Prepare a large set of potential design points $\mathcal{C}$\;
Choose the set of basis functions of the differential equations\;
Derive the formula of the necessary derivative functions of the GP predictor\;
Generate the initial design $\mathcal{D}$ by random sampling $N_0$ design points from the potential design points\;
Collect the data based on the initial design\;
Initialization: set $\bm \beta_c={\bf 1}$, $\bm \beta_o={\bf 0}$, the current sample size $n=N_0$\;
\While{$\frac{\|\bm \beta_c-\bm \beta_o\|_2}{\|\bm \beta_c\|_2}\geq Tol$ and $n\leq N_{\max}$}
{
Update $\bm \beta_o\leftarrow \bm \beta_c$\;
Based on the current observations, construct the basis functions at the newly selected design points, and form $\mathbf {M}_n$\;
Use forward stepwise regression and BIC criterion to fit the regression model such as \eqref{eq:reg_setup}. Obtain $\hat{\sigma}^2$ and estimate linear coefficients $\bm \beta_c$ (If a basis function is not selected into the stepwise regression, set the corresponding coefficient to zero.)\;
Fit the GP surrogate model(s) with the currently collected data and compute the leave-one-out cross-validation error $\hat{\tau}_{cv}^2$\;
Using the GP predictor(s) and the derivatives (with estimated parameters), calculate the values of the basis functions $\bm m(\bm x)$ at the potential design points\;
\For {$j=1,\ldots, B$}{
Update $U_S$ and $U_D$\;
Compute the ACDS design criterion \eqref{eq:acds} for each potential design point in $\mathcal{C}$\;
Select the design point with largest ACDS criterion into $\mathcal{D}$ and remove it from $\mathcal{C}$\;
Update $(\mathbf {M}_{n+j-1}^\top \mathbf {M}_{n+j-1})^{-1}$ to $(\mathbf {M}_{n+j}^\top \mathbf {M}_{n+j})^{-1}$\;
}
Collect the data for the newly $B$ selected design points\;
Update $n\leftarrow n+B$.
}
\caption{Gaussian Process assisted active learning of physical laws. \label{alg:actlearn}}
\end{algorithm}

\section{Numerical Case Studies}
\label{sec:simu}

In this section, we use several simulation case studies to demonstrate the performances of the proposed active learning approach.
We generate the data using a known PDE or ODE system and then use an active learning method to identify the differential equations and compare them with the true ones.
We compare the active learning with the ACDS criterion with maximin space-filling design.
Although both final designs are sequentially constructed, maximin space-filling design does not need GP surrogate model since it is model-free.

We measure the performance of different methods on three aspects: variable selection accuracy, parameter estimation accuracy, and the size of the total design points denoted as $N$.
On variable selection, we consider both the number of false-positive (FP) and false-negative (FN) cases.
In the FP case, the variable selection method mistakenly identifies some terms as significant, but they are not included in the underlying equations.
The opposite case is when some terms contained by the true equations are missed by the variable selection method.
We define the total number of falsely identified terms by $\gamma=FP+FN$ to account for both cases.
To evaluate the parameter estimation accuracy, $l_2$ loss is considered as follows
\[
l_2(\bm \beta)=\|\hat{\bm \beta}-\bm \beta_{true}\|_2
\]
where $\|\cdot\|_2$ stands for the $l_2$ norm, $\hat{\bm \beta}$ is estimated parameter values and $\bm \beta_{true}$ is true parameter values.

\subsection{An ODE system}\label{subsection:ODE}

Consider the two-dimensional ODE system
\[\left\{
\begin{aligned}
\frac{dy_1}{dx} =-0.5 y_1+2 y_2\\
\frac{dy_2}{dx} =-2 y_1-0.5 y_2,
\end{aligned}
\right.
\]
as the true underlying differential equations.
We set the initial condition to be $(y_1,y_2)|_{x=0}=(2,0)$.
The system is solved by the MATLAB ODE solver \verb|ode45|. 
Independent noise $\epsilon$ is added to $dy_1/dx$ and $dy_2/dx$.
The ranges of $dy_1/dx$ and $dy_2/dx$ (without noise) are $[-3.052,1.392]$ and $[-4.000,2.061]$, and the standard deviations are $0.544$ and $0.519$.
The variance of noise are set to be $\sigma^2=0.2^2, 0.5^2, 0.8^2$.

The linear regression models we use to learn the ODE system are
\begin{align*}
\frac{dy_1}{dx}&=\bm f(y_1,y_2)^\top \bm \beta_1\\
\frac{dy_2}{dx}&=\bm f(y_1,y_2)^\top \bm \beta_2,
\end{align*}
where $\bm f(y_1,y_2)$ is the vector of candidate basis functions that are monomials of $y_1$ and $y_2$ to the fifth degree, and $\bm \beta_1$ and $\bm \beta_2$ are regression coefficients to be estimated.
The potential design points in $\mathcal{C}$ are $3000$ equally spaced points in the interval $[0, T]$ with $T=30$.
The initial design contains $N_0=16$ randomly selected design points from $\mathcal{C}$.
The batch size is $B=16$ in each iteration of active learning.
Table \ref{tab:identified_ODE} shows the identified ODE system with the 95\% confidence intervals of the coefficient parameters (in the parenthesis) from a single simulation for each $\sigma^2$ setting.

\begin{figure}[ht]
\centering
\includegraphics[width=\linewidth]{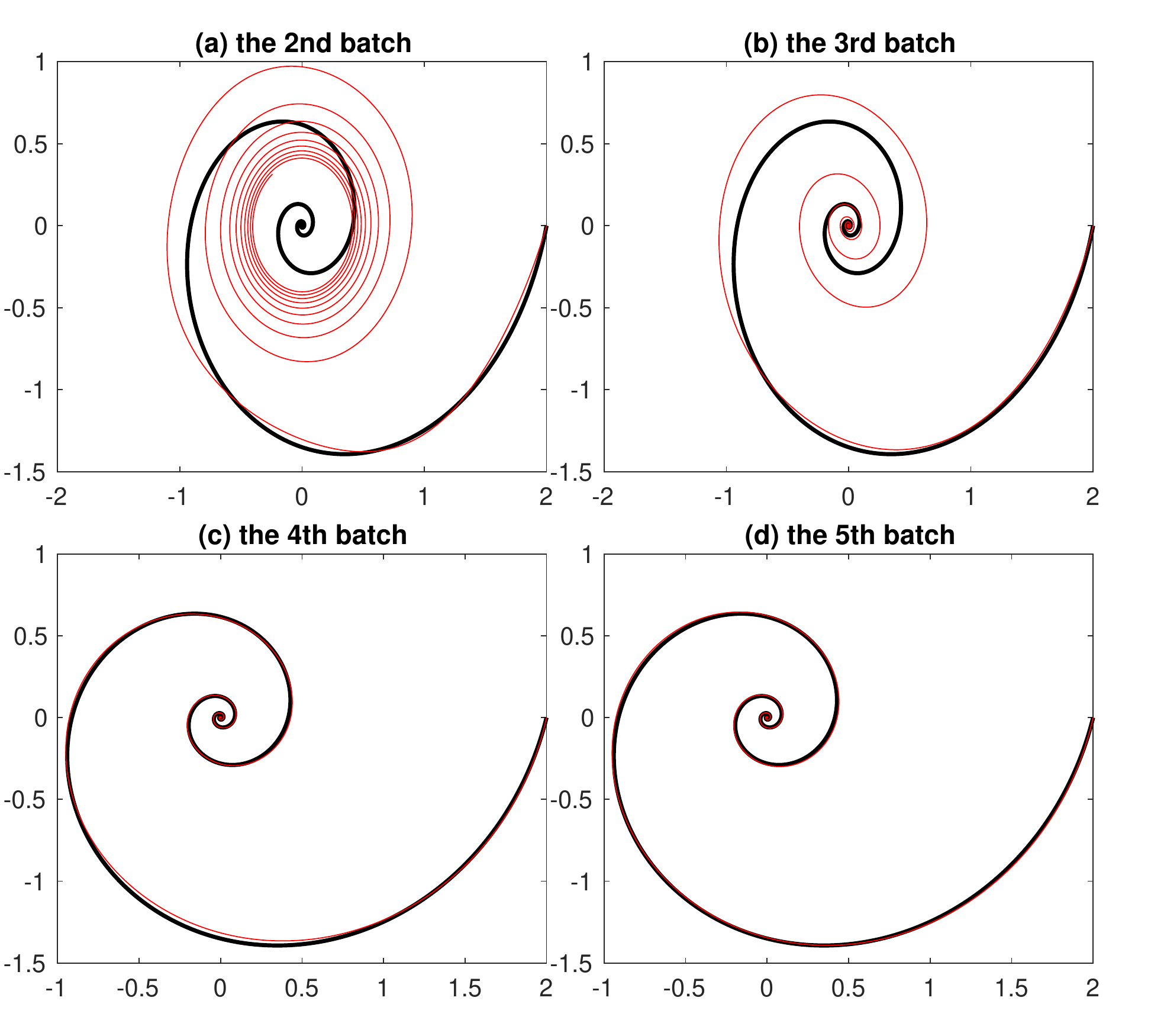}
\caption{Four snapshots of the solution of the estimated ODE system. Red line represents solution of the estimated equation and black line represents true solution.
\label{fig:ODE}}
\end{figure}

In Figure \ref{fig:ODE}, the progress of the proposed active learning with the ACDS criterion is shown.
In this simulation we set $\sigma^2=0.5^2$.
The solution of the estimated ODE system is compared with the true solution when the sequential design reaches the size of 48, 64, 80, and 96, and it becomes closer to the true solution path as more data are collected.
Eventually, the two solution paths almost overlap each other, indicating the accuracy of the proposed active learning approach.
This case study is also shown in \cite{zhang2018robust}, in which $N=200$ design points are used in a non-sequential design of the experiment.
Comparing Figure \ref{fig:ODE} with Figure 2 in \cite{zhang2018robust} with $\sigma^2=0.5^2$, we can see that the active learning method with ACDS criterion performs equally well as the threshold sparse Bayesian regression proposed by \cite{zhang2018robust} in terms of accuracy of model estimation.
But the active learning method uses only about half of the data, and the forward stepwise regression is a much simpler variable selection technique than that of \cite{zhang2018robust}.

Table \ref{tab:ODE1} and Table \ref{tab:ODE2} compare ACDS active learning with sequential maximin space-filling deisgn.
Both show the mean and standard deviation of the performance measures from the 50 simulations.
In Table \ref{tab:ODE1}, the sequential procedure is terminated when convergence is reached, i.e., $||\bm \beta_c-\bm \beta_o||/||\bm \beta_c||<Tol$ and $Tol=10^{-2}$.
We can see from Table \ref{tab:ODE1} that the ACDS criterion outperforms the maximin space-filling design in terms of the variable selection and parameter estimation, although it uses a larger sample to converge.
Both methods become worse when the variance of the noise becomes larger, as expected.
In Table \ref{tab:ODE2}, we fix sample size to be $N=112$, and thus the convergence condition is not necessarily always guaranteed for either of the two methods.
It is obvious that given the same amount of data the ACDS returns a much more accurate identified ODE system than maximin space-filling design.

\begin{table}[ht]
\centering
\caption{ODE System: identified ODE's and 95\% C.I. of parameters.}
\label{tab:identified_ODE}
\begin{tabular}{ll}\hline
true system &$dy_1$/$dx =-0.5y_1+2y_2$\\
&$dy_2$/$dx =-2y_1-0.5y_2$\\\hline
$\sigma^2=0.2^2$ &$dy_1$/$dx =-0.499(\pm 0.0323)y_1+1.986(\pm 0.0487)y_2$\\
&$dy_2$/$dx =-1.974(\pm 0.0306)y_1-0.517(\pm 0.0461)y_2$\\\hline
$\sigma^2=0.5^2$ &$dy_1$/$dx =-0.540(\pm 0.0479)y_1+2.070(\pm 0.0638)y_2$\\
&$dy_2$/$dx =-1.967(\pm 0.0482)y_1-0.505(\pm 0.0642)y_2$\\\hline
$\sigma^2=0.8^2$ &$dy_1$/$dx =-0.496(\pm 0.0714)y_1+1.944(\pm 0.0928)y_2$\\
&$dy_2$/$dx =-2.044(\pm 0.0734)y_1-0.509(\pm 0.0955)y_2$\\\hline
\end{tabular}
\end{table}

\begin{table}[ht]
\centering
\caption{Comparison of two methods when both reach convergence for a pre-set ODE system.}
\label{tab:ODE1}
\begin{tabular}{lllllll}\hline
&&\multicolumn{2}{c}{ACDS}&&\multicolumn{2}{c}{maximin space-filling}\\\cline{3-4} \cline{6-7}
&&\multicolumn{2}{c}{\small mean (std.)}&&\multicolumn{2}{c}{\small mean (std.)}\\\hline
&$\gamma$&\multicolumn{2}{c}{0.360 (0.598)}&&\multicolumn{2}{c}{0.560 (1.033)}\\
$\sigma^2=0.2^2$&$l_2(\bm \beta)$&\multicolumn{2}{c}{0.107 (0.099)}&&\multicolumn{2}{c}{0.306 (0.438)}\\
&$N$&\multicolumn{2}{c}{121.280 (49.033)}&&\multicolumn{2}{c}{93.440 (26.726)  }\\\hline
&$\gamma$&\multicolumn{2}{c}{0.400 (0.670)}&&\multicolumn{2}{c}{1.220 (1.556)}\\
$\sigma^2=0.5^2$&$l_2(\bm \beta)$&\multicolumn{2}{c}{0.171 (0.156)}&&\multicolumn{2}{c}{1.055 (2.406)}\\
&$N$&\multicolumn{2}{c}{169.920 (59.944)}&&\multicolumn{2}{c}{106.240 (29.584)}\\\hline
&$\gamma$&\multicolumn{2}{c}{0.580 (0.928)}&&\multicolumn{2}{c}{2.400 (1.195)}\\
$\sigma^2=0.8^2$&$l_2(\bm \beta)$&\multicolumn{2}{c}{0.271 (0.206)}&&\multicolumn{2}{c}{1.162 (0.747)}\\
&$N$&\multicolumn{2}{c}{265.280 (84.793)}&&\multicolumn{2}{c}{122.880 (48.211)}\\\hline
\end{tabular}
\end{table}

\begin{table}[htb]
\centering
\caption{Comparison of two methods with the same fixed sample size for a pre-set ODE system.}
\label{tab:ODE2}
\begin{tabular}{lllllll}\hline
&&\multicolumn{2}{c}{ACDS}&&\multicolumn{2}{c}{maximin space-filling}\\\cline{3-4} \cline{6-7}
&&\multicolumn{2}{c}{\small mean (std.)}&&\multicolumn{2}{c}{\small mean (std.)}\\\hline
&$\gamma$&\multicolumn{2}{c}{0.440 (0.675)}&&\multicolumn{2}{c}{0.440 (0.812)}\\
$\sigma^2=0.2^2$&$l_2(\bm \beta)$&\multicolumn{2}{c}{0.100 (0.072)}&&\multicolumn{2}{c}{0.218 (0.163)}\\
&$N$&\multicolumn{2}{c}{112  }&&\multicolumn{2}{c}{112 }\\\hline
&$\gamma$&\multicolumn{2}{c}{0.620 (1.067)}&&\multicolumn{2}{c}{1.560 (1.445)}\\
$\sigma^2=0.5^2$&$l_2(\bm \beta)$&\multicolumn{2}{c}{0.262 (0.209)}&&\multicolumn{2}{c}{0.769 (0.772)}\\
&$N$&\multicolumn{2}{c}{112}&&\multicolumn{2}{c}{112}\\\hline
&$\gamma$&\multicolumn{2}{c}{1.260 (1.412)}&&\multicolumn{2}{c}{2.320 (1.421)}\\
$\sigma^2=0.8^2$&$l_2(\bm \beta)$&\multicolumn{2}{c}{0.501 (0.386)}&&\multicolumn{2}{c}{1.299 (0.958)}\\
&$N$&\multicolumn{2}{c}{112}&&\multicolumn{2}{c}{112}\\\hline
\end{tabular}
\end{table}

\subsection{An ODE system with random coefficients}

To show the robustness of ACDS, we modify the previous ODE example into a more challenging case.
Consider the two-dimensional ODE system
\[
\left\{
\begin{aligned}
\frac{dy_1}{dx} =-a y_1+b y_2\\
\frac{dy_2}{dx} =-b y_1-a y_2,
\end{aligned}
\right.
\]
where $a$ and $b$ are randomly sampled from $\text{Uniform}[0.5,1.5]$ and $\text{Uniform}[2,3]$.
The variance of the noise is set to be $\sigma^2=0.4^2, 0.6^2, 0.8^2$.
For a given $\sigma^2$, we run one simulation for a pair of randomly sampled $(a,b)$ values.
We simulate 50 times for each $\sigma^2$ setting and show the mean and standard deviation of $\gamma$ and $l_2(\bm \beta)$ in Table \ref{tab:ODE3}.
We use the same settings for active learning as in the previous case.
The sample size for both methods is fixed at $N=112$, and thus the convergence condition is not necessarily always reached.
It is obvious that given the same amount of data the ACDS returns a much more accurate identified ODE system and this result is consistent when the underlying ODE systems are varied.
\begin{table}[ht]
\centering
\caption{Comparison of two methods for ODE System with random coefficients.}
\label{tab:ODE3}
\begin{tabular}{lllllll}\hline
&&\multicolumn{2}{c}{ACDS}&&\multicolumn{2}{c}{maximin space-filling}\\\cline{3-4} \cline{6-7}
&&\multicolumn{2}{c}{\small mean (std.)}&&\multicolumn{2}{c}{\small mean (std.)}\\\hline
&$\gamma$&\multicolumn{2}{c}{0.620 (0.855)}&&\multicolumn{2}{c}{1.160 (1.490)}\\
$\sigma^2=0.4^2$&$l_2(\bm \beta)$&\multicolumn{2}{c}{0.380 (0.470)}&&\multicolumn{2}{c}{1.221 (1.437)}\\
&$N$&\multicolumn{2}{c}{112  }&&\multicolumn{2}{c}{112 }\\\hline
&$\gamma$&\multicolumn{2}{c}{0.760 (1.135)}&&\multicolumn{2}{c}{1.800 (1.852)}\\
$\sigma^2=0.6^2$&$l_2(\bm \beta)$&\multicolumn{2}{c}{0.614 (0.744)}&&\multicolumn{2}{c}{2.065 (2.153)}\\
&$N$&\multicolumn{2}{c}{112}&&\multicolumn{2}{c}{112}\\\hline
&$\gamma$&\multicolumn{2}{c}{1.280 (1.565)}&&\multicolumn{2}{c}{2.260 (1.440)}\\
$\sigma^2=0.8^2$&$l_2(\bm \beta)$&\multicolumn{2}{c}{1.093 (1.233)}&&\multicolumn{2}{c}{2.332 (1.693)}\\
&$N$&\multicolumn{2}{c}{112}&&\multicolumn{2}{c}{112}\\\hline
\end{tabular}
\end{table}

\subsection{Bass model with random coefficients}
The Bass model is a simple differential equation that is widely used in marketing research.
It describes the process that new products get adopted by a mass population.
Consider the one-dimensional Bass model
\[
\frac{dF}{dt} =(1-F)(p+qF)
\]
as the true underlying differential equation.
The coefficient $p$ is called the coefficient of innovation, external influence, or advertising effect, which has a typical range between $[0,0.03]$.
The coefficient $q$ is called the coefficient of imitation, internal influence, or word-of-mouth effect, with a typical range between $[0.3,0.5]$ \citep{mahajan1995diffusion}.
One nice feature of this Bass model is that it has a tractable solution,
\[
F(t) = \frac{1-e^{-(p+q)t}}{1+\frac{q}{p}e^{-(p+q)t}}.
\]
So we can simply generate the observational data from this solution instead of solving the original differential equation.

The candidate basis functions for active learning are polynomials of $F(t)$ to the fifth degree.
The potential design points are $3000$ equally spaced points in the time interval $[0, T]$ with $T=30$.
The initial design contains $N_0=16$ randomly selected design points from the potential design points.
The batch size is $B=16$.
The coefficients $p$ and $q$ are randomly generated from uniform distributions in the range of $[0,0.03]$ and $[0.3,0.5]$.
In Table \ref{tab:Bass}, we compare the proposed ACDS active learning with maximin space-filling designs.
For both methods, we let the active learning procedure run long enough until the convergence condition $||\bm \beta_c-\bm \beta_o||/||\bm \beta_c||<Tol$ and $Tol=10^{-2}$ is reached.
The noise is added directly to $F(t)$ observations, and we set $\sigma^2=0.01^2, 0.02^2, 0.04^2$.
For each setting of $\sigma^2$, we run 50 simulations.
The range of $dF/dt$ (without noise) is $[0,0.091]$.
As similarly in the previous two examples, the proposed ACDS active learning is superior to the space-filling design.

\begin{table}[ht]
\centering
\caption{Comparison of two methods on the Bass model.}
\label{tab:Bass}
\begin{tabular}{lllllll}\hline
&&\multicolumn{2}{c}{ACDS}&&\multicolumn{2}{c}{maximin space-filling}\\\cline{3-4} \cline{6-7}
&&\multicolumn{2}{c}{\small mean (std.)}&&\multicolumn{2}{c}{\small mean (std.)}\\\hline
&$\gamma$&\multicolumn{2}{c}{0.420 (0.859)}&&\multicolumn{2}{c}{0.620 (1.105)}\\
$\sigma^2=0.01^2$&$l_2(\bm \beta)$&\multicolumn{2}{c}{0.068 (0.168)}&&\multicolumn{2}{c}{0.119 (0.260)}\\
&$N$&\multicolumn{2}{c}{87.360 (30.531) }&&\multicolumn{2}{c}{90.240 (55.025)}\\\hline
&$\gamma$&\multicolumn{2}{c}{0.740 (1.065)}&&\multicolumn{2}{c}{1.020 (1.134)}\\
$\sigma^2=0.02^2$&$l_2(\bm \beta)$&\multicolumn{2}{c}{0.131 (0.261)}&&\multicolumn{2}{c}{0.199 (0.307)}\\
&$N$&\multicolumn{2}{c}{139.520 (53.609)}&&\multicolumn{2}{c}{123.200 (56.661)}\\\hline
&$\gamma$&\multicolumn{2}{c}{1.300 (1.111)}&&\multicolumn{2}{c}{1.660 (1.136)}\\
$\sigma^2=0.04^2$&$l_2(\bm \beta)$&\multicolumn{2}{c}{0.271 (0.326)}&&\multicolumn{2}{c}{0.375 (0.435)}\\
&$N$&\multicolumn{2}{c}{182.720 (87.522)}&&\multicolumn{2}{c}{164.480 (72.356)}\\\hline
\end{tabular}
\end{table}

\subsection{Burgers' equation}
Burgers' equation is one of the most important PDEs applied in various areas of physics, such as fluid mechanics, nonlinear acoustics, gas dynamics, and traffic flow.
It can be derived from the Navier-Stokes equation for the velocity by dropping the pressure gradient term.
For the one-dimensional space, i.e., $x\in \mathbb{R}^1$, the Burgers' equation is
\begin{equation}
\label{eq:burgers'}
u_t+\lambda_1uu_x-\lambda_2u_{xx}=0,
\end{equation}
where the parameters $(\lambda_1,\lambda_2)$ are set to be $(1, -0.01)$.
The initial condition is chosen as $u_0(x)=2\exp(-15(x-6)^2)+1.5\exp(-15(x + 1)^2)+\exp(-25(x + 5)^2)$.
For Burgers' equation, theoretically, $x\in (-\infty,\infty)$, and thus there is no boundary condition.
But to solve it numerically, we need to restrict $x$ in a bounded domain.
To generate the observation data, we solve the Burgers' equation by finite difference with time step $\delta_t =0.001$ and space step $\delta_x=0.0025$ in the region $t\in[0,1]$ and $x\in[0,10]$.
Independent noise $\epsilon$ is added to $u_t$ in \eqref{eq:burgers'}, and its variance  $\sigma^2$ is set to be $0.2^2$, $0.4^2$ and $0.8^2$.
The range of $u_t$ (without noise) is $[-5.8,17.3]$ and the standard deviation is $1.43$.
We use the same candidate basis functions as in \cite{schaeffer2017learning}, which are
\[
\{1,u,u^2,u^3,u_x,u_x^2,u_x^3,uu_x,u^2u_x,uu_x^2,u_{xx},u_{xx}^2,u_{xx}^3,uu_{xx},u^2u_{xx},uu_{xx}^2,u_xu_{xx},u_{x}^2u_{xx},u_xu_{xx}^2,uu_xu_{xx}\}.
\]
As explained in Section 2, to select the significant terms from these candidates, we can regress $u_t$ against these basis functions that do not involve time $t$.
Therefore, we only need to collect the necessary observations at a certain time point, $t=t_s$.
Here we choose $t_s=0.1$ (actually, we can choose any time) and collect all the necessary of observations of $u_t$, $u$, and the other basis functions at $t_s=0.1$.

The set $\mathcal{C}$ of potential design points contains $4,000$ equally spaced points in $[0,10]$.
The initial design contains $N_0=5$ randomly chosen design points from $\mathcal{C}$.
The batch size is $B=10$.
In Table \ref{tab:Burgers1}, we show the identified equation with 95\% confidence intervals of the coefficients (in the parenthesis) from a single simulation for each setting of $\sigma^2$.
In Table \ref{tab:Burgers2}, we compare active learning with ACDS and maximin space-filling design.
We first perform the active learning approach and discover that on average the procedure converges when $N$ reaches approximately 70, 72, 90, and 97 for $\sigma^2=0.2^2, 0.4^2$, and $0.8^2$.
Thus, we fix the space-filling design with size $N=70, 72, 90$, and $97$.
We run the simulation 100 times and compare the mean and standard deviation of $\gamma$ and $l_2(\bm \beta)$.
For this PDE case study, the proposed approach still outperforms the random design in terms of accuracy.

\begin{table}[ht]
\centering
\caption{Burgers' Equation: true equation and the identified equation for different $\sigma^2$ value.}\label{tab:Burgers1}
\begin{tabular}{ll}\hline
true system &$u_t+uu_x-0.01u_{xx}=0$\\\hline
$\sigma^2=0.2^2$ &$u_t+0.9982(\pm 0.0093)uu_x-0.0108(\pm 0.0009)u_{xx}=0$\\\hline
$\sigma^2=0.4^2$ &$u_t+0.9928(\pm 0.0103)uu_x-0.0092(\pm 0.0009)u_{xx}=0$\\\hline
$\sigma^2=0.8^2$ &$u_t+0.9949(\pm 0.0164)uu_x-0.0114(\pm 0.0015)u_{xx}=0$\\\hline
\end{tabular}
\end{table}

\begin{table}[ht]
\centering
\caption{Comparison of ACDS active learning and maximin space-filling for Burgers' equation.}
\label{tab:Burgers2}
\begin{tabular}{lllllll}\hline
&&\multicolumn{2}{c}{ACDS}&&\multicolumn{2}{c}{maximin space-filling}\\\cline{3-4} \cline{6-7}
&&\multicolumn{2}{c}{\small mean (std.)}&&\multicolumn{2}{c}{\small mean (std.)}\\\hline
&$\gamma$&\multicolumn{2}{c}{0.330 (0.668)}&&\multicolumn{2}{c}{0.840 (1.051)}\\
$\sigma^2=0.2^2$&$l_2(\bm \beta)$&\multicolumn{2}{c}{0.016 (0.031)}&&\multicolumn{2}{c}{0.123 (0.406)}\\
&$N$&\multicolumn{2}{c}{68.400 (24.380)}&&\multicolumn{2}{c}{70 }\\\hline
&$\gamma$&\multicolumn{2}{c}{0.540 (0.834)}&&\multicolumn{2}{c}{1.310 (0.961)}\\
$\sigma^2=0.4^2$&$l_2(\bm \beta)$&\multicolumn{2}{c}{0.028 (0.041)}&&\multicolumn{2}{c}{0.220 (0.484)}\\
&$N$&\multicolumn{2}{c}{72.100 (27.535)}&&\multicolumn{2}{c}{72 }\\\hline
&$\gamma$&\multicolumn{2}{c}{0.930 (0.946)}&&\multicolumn{2}{c}{1.240 (0.726)}\\
$\sigma^2=0.8^2$&$l_2(\bm \beta)$&\multicolumn{2}{c}{0.082 (0.183)}&&\multicolumn{2}{c}{0.214 (0.457)}\\
&$N$&\multicolumn{2}{c}{97.100 (52.229)}&&\multicolumn{2}{c}{97}\\\hline
\end{tabular}
\end{table}

\section{A Real Example: Air Pollution Monitoring}
\label{sec:air}

To motivate the use of our method in a practical application, we consider the problem of air pollution monitoring.
In particular, we aim to compute sequential optimal sensor placement for systems described by advection and diffusion equation \citep{egan1972numerical,scheff1992source},
\[
\frac{\partial c}{\partial t} = \nabla \cdot (D\nabla) - \nabla \cdot (\mathbf{v}c),
\]
where $c(\bm x, t)$ is the concentration of air pollution at location $\bm x$ and time $t$.
Constant $D$ is the diffusion coefficient.
The higher the diffusivity, the faster the pollutant diffuses into the air.
The vector $\mathbf{v}$ is the velocity field of air movement.
The experimenter allocates limited sensor to monitor air pollution and then identifies the diffusion coefficient $D$ \citep{yu2018scalable,cheng2011modeling} .
Unfortunately, the concentration of air pollution is difficult to assess experimentally due to the limited number of monitors \citep{cheng2011modeling}.

\begin{figure}
\centering
\includegraphics[width=\linewidth]{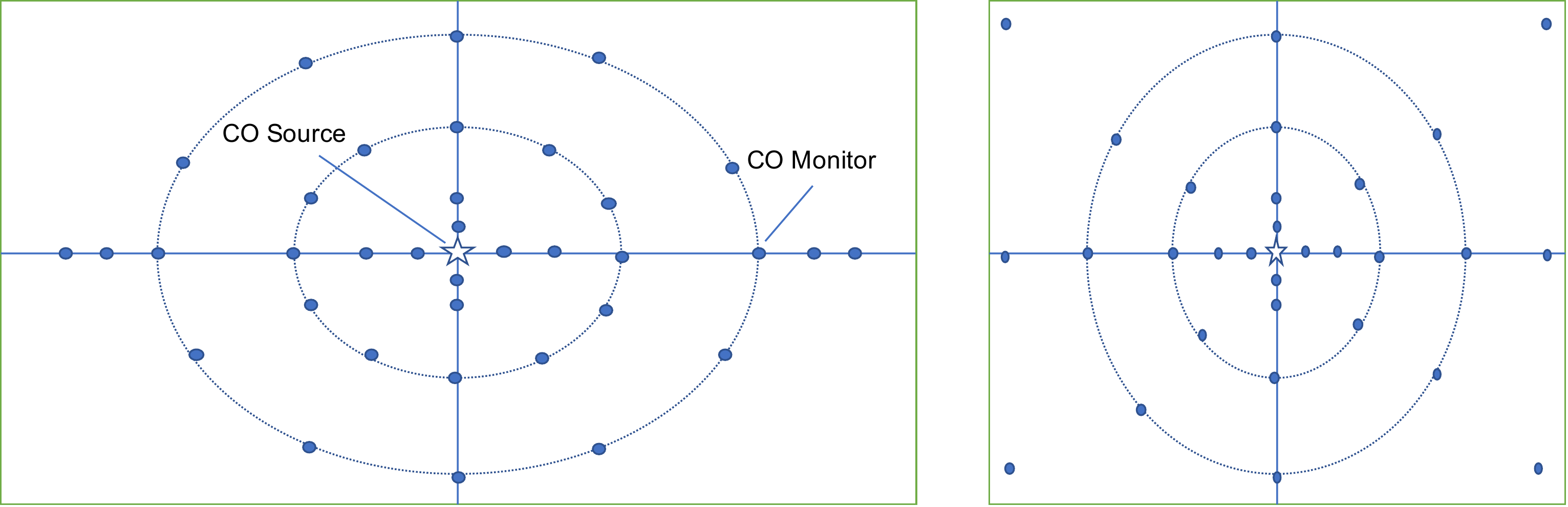}
\caption{Air pollution: View of CO monitor placement for two different residential houses. The unfilled star presents the CO point source located at the center of room; the solid dots represent the CO monitors.}
\label{sensor}
\end{figure}

\begin{figure}
\centering
\begin{subfigure}[b]{0.7\textwidth}
\includegraphics[width=\linewidth]{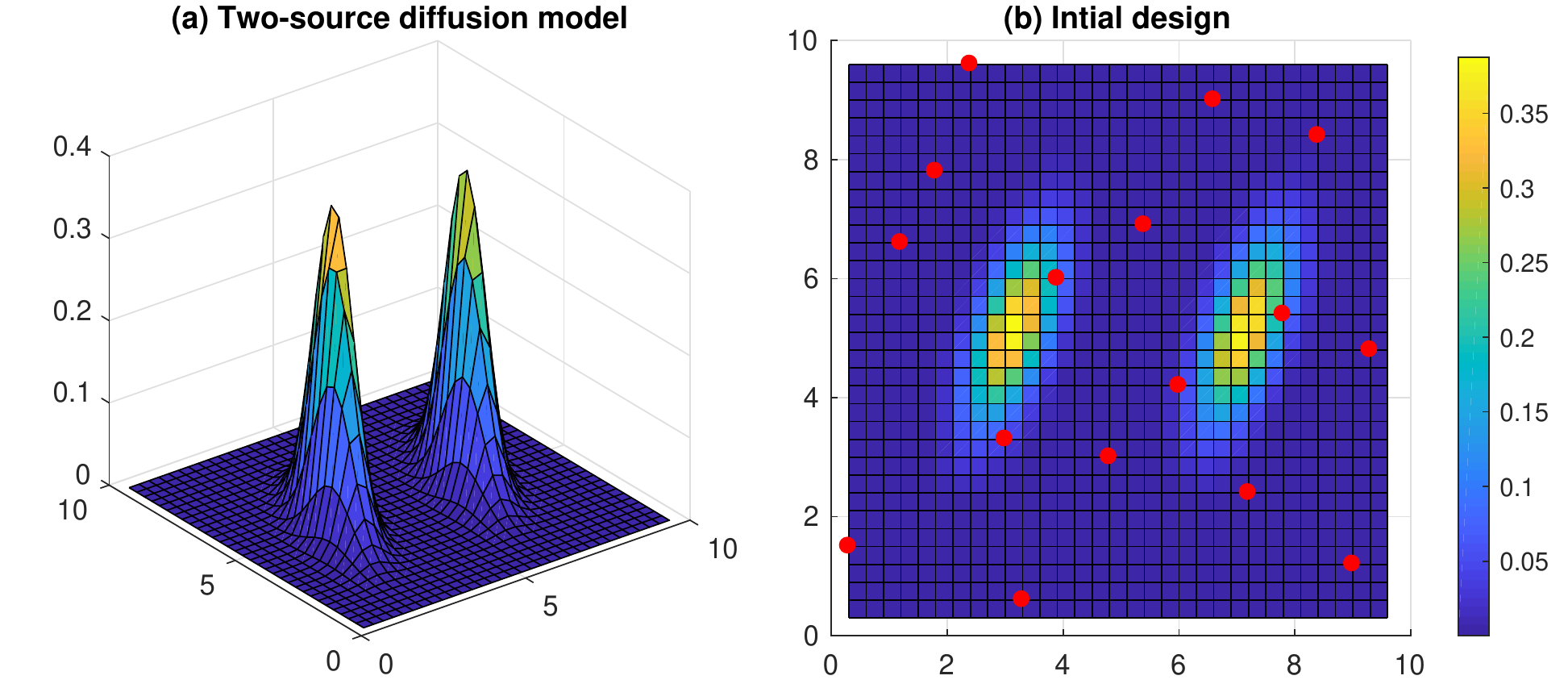}
\caption*{(I) Reference}
\end{subfigure}
 ~
\begin{subfigure}[b]{0.7\textwidth}
\includegraphics[width=\linewidth]{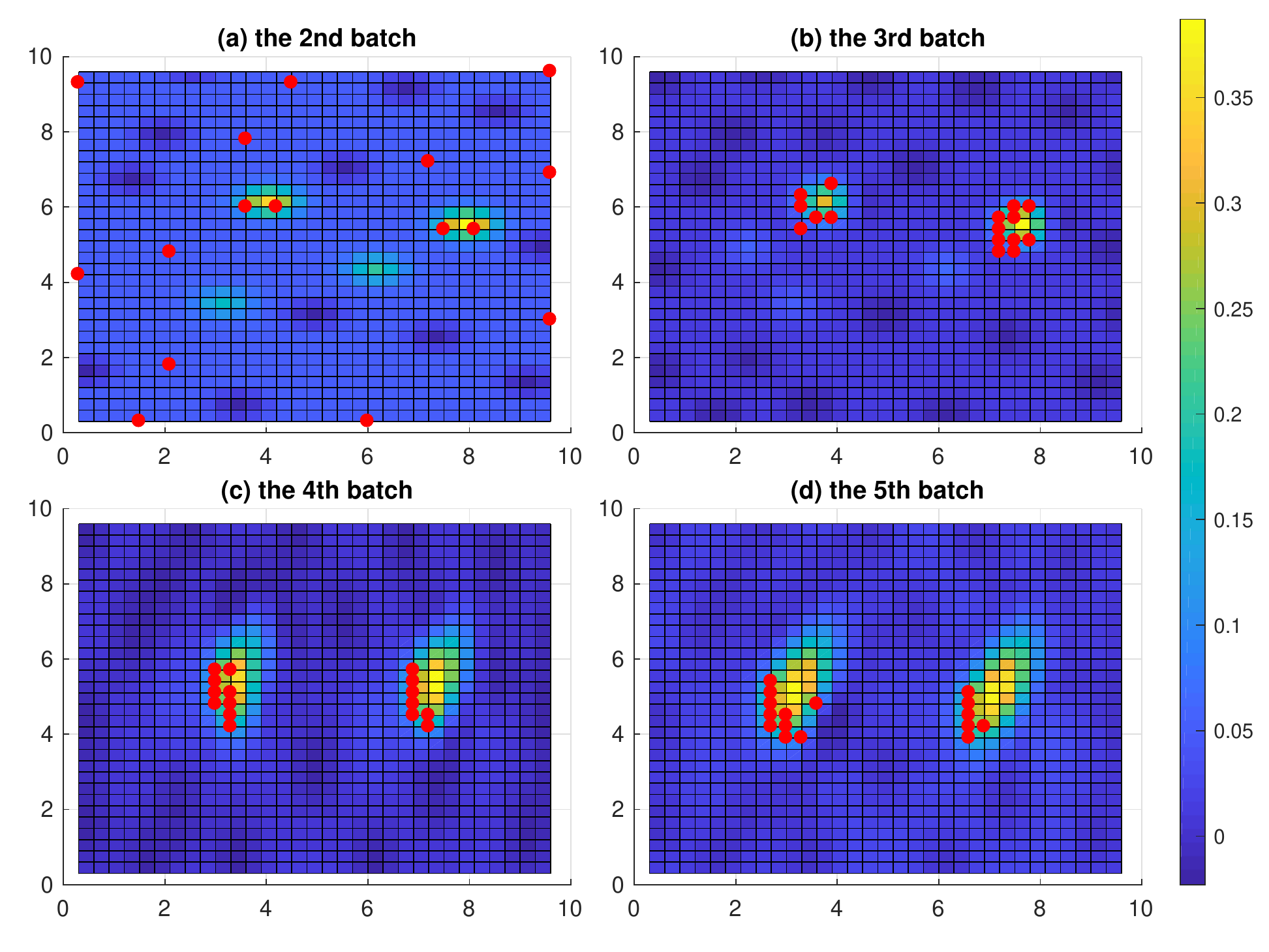}
\caption*{(II) ACDS}
\end{subfigure}
\caption{Air pollution: four snapshots of sequentially added batch of design points for $\sigma^2=0.2^2$.}
\label{fig:heat1}
\end{figure}

The conventional practice is that the experimenter usually allocates these limited sensors in such a configuration that each monitor surrounds the air pollution source at different radial distances and angles \citep{cheng2011modeling}.
Figure \ref{sensor} illustrates the placement of sensors for two residential houses with different shapes.
CO monitors are placed on the circles centering at the CO source.
This pattern has two restrictions. 
First, such placement can only be implemented when the experimenter is aware of the pollution source position. 
But it is not always the case in practical applications.
Second, this experimental setting assumes that the pollution concentration is spatially homogeneous within the room.
Such placement of sensors in Figure \ref{sensor} can not be used in a more general setting.
For instance, we consider the Two-Source Advection-Diffusion Model \citep{scheff1992source} shown in Figure \ref{fig:heat1} (I)-(a).
In this scenario, the concentration of air pollution is no longer spatially homogeneous within an indoor space.

In this section, we consider an experiment on air pollution monitoring.
Besides the maximin space-filling design, we also compare it with the D-optimality criterion.
To implement the D-optimality in the active learning, we only need to use \eqref{criterion} rather than \eqref{eq:acds} and set $\alpha_1=0$ throughout Algorithm \ref{alg:actlearn}. 
Other components of Algorithm \ref{alg:actlearn} are still the same.
We make the following assumptions on the experiment.
\begin{enumerate}
\item The experimenter is not aware of the location of the source of pollution, but only the initial condition of the pollution concentration.
The experimenter hopes to collect data and identify what the diffusion process is, i.e., the function format, as well as the parameters.
\item The underlying pollution concentration is spatially heterogeneous, and we further assume the true underlying 2-D diffusion equation
\[
\frac{\partial c}{\partial t} = c_{xx}+c_{yy},
\]
with 0 boundary condition and initial condition shown in Figure \ref{fig:heat1} (I)-(a).
This initial condition is the sum of two multivariate normal probability density functions with two different means,  $\mu_1=(3,5)$ and $\mu_2=(7,5)$, and a common covariance matrix $\Sigma=[0.25,0.3; 0.3,1]$.
\item The experimenter can either move the sensors to new locations in a negligible short time or repeat the diffusion process multiple times with the exact initial condition so that each new batch of data p are collected at the same time point $t=t_s$.
\end{enumerate}

\begin{table}[ht]
\centering
\caption{Basis functions for air pollution monitoring problem.}\label{tab:heat1}
\begin{tabular}{ll}\hline
first order & $c,c_x,c_y,c_{xx},c_{yy},c_{xy}$\\\hline
second order & $c^2,cc_x, cc_y, c_x^2, c_y^2, c_yc_x, c_{xx}c, c_{xx}c_x, c_{xx}c_y, c_{yy}c, c_{yy}c_x, c_{yy}c_y, c_{xy}c, c_{xy}c_x, c_{xy}c_y,$\\
&$c_{xx}^2, c_{xx}c_{yy}, c_{xx}c_{xy}, c_{xy}^2, c_{yy}^2, c_{yy}c_{xy}$\\\hline
\end{tabular}
\end{table}

To implement the proposed active learning approach, we first construct the candidate basis functions in Table \ref{tab:heat1}.
Set the time $t_s=0.0005$ and collect all the data at the spatial locations.
The potential design points in $\mathcal{C}$ are $32\times 32$ equally spaced grid points in $[0,10]\times[0,10]$ domain.
Let us assume that there are 16 sensors used.
The initial design is a Latin Hypercube space-filling design chosen from $\mathcal{C}$ and the batch size is $B=16$ as shown in Figure \ref{fig:heat1} (II)-(b).
In this practical application, we decide to collect 5 batches of data, and thus the total number of collected data is $N_{\max}=16\times 5=80$.
In Table \ref{tab:heat2}, we show the identified equation with 95\% confidence interval of parameters from a single simulation for different setting of $\sigma^2$.
We also show the progress of the addition of the design points in the active learning process in Figure \ref{fig:heat1} (II).
In each of the sub-figure, the red dots represent the newly added batch of design points.
The heatmap is generated by the fitted surrogate model of the available observations.
Table \ref{tab:heat2} and Figure \ref{fig:heat1} show that ACDS can identify not only the diffusion coefficient accurately but also the position of two sources correctly.

\begin{table}[ht]
\centering
\caption{True PDE and identified PDE for air pollution monitoring problem.}\label{tab:heat2}
\begin{tabular}{ll}\hline
true system &$c_t=c_{xx}+c_{yy}$\\\hline
$\sigma^2=0.2^2$ &$c_t=0.9988(\pm 0.1319)c_{xx}+1.0098(\pm 0.0287)c_{yy}$\\\hline
$\sigma^2=0.4^2$ &$c_t=1.0544(\pm 0.2734)c_{xx}+1.0145(\pm 0.0665)c_{yy}$\\\hline
$\sigma^2=0.8^2$ &$c_t=0.8421(\pm 0.3112)c_{xx}+0.9951(\pm 0.0770)c_{yy}$\\\hline
\end{tabular}
\end{table}

\begin{table}[ht]
\centering
\caption{Comparison of three methods for air pollution monitoring problem.}
\label{tab:heat3}
\begin{tabular}{llllllllll}\hline
&&\multicolumn{2}{c}{ACDS}&&\multicolumn{2}{c}{D-optimal}&&\multicolumn{2}{c}{maximin space-filling}\\
\cline{3-4} \cline{6-7}\cline{9-10}
&&\multicolumn{2}{c}{\small mean (std.)}&&\multicolumn{2}{c}{\small mean (std.)}&&\multicolumn{2}{c}{\small mean (std.)}\\\hline
&$\gamma$&\multicolumn{2}{c}{0.440 (0.577)}&&\multicolumn{2}{c}{0.380 (0.635)}&&\multicolumn{2}{c}{0.780 (0.975)}\\
$\sigma^2=0.2^2$&$l_2(\bm \beta)$&\multicolumn{2}{c}{0.392 (0.495)}&&\multicolumn{2}{c}{0.500 (0.960)}&&\multicolumn{2}{c}{1.738 (2.947)}\\\hline
&$\gamma$&\multicolumn{2}{c}{0.940 (1.114)}&&\multicolumn{2}{c}{0.700 (0.909)}&&\multicolumn{2}{c}{1.420 (0.835)}\\
$\sigma^2=0.4^2$&$l_2(\bm \beta)$&\multicolumn{2}{c}{1.620 (2.029)}&&\multicolumn{2}{c}{1.633 (2.298)}&&\multicolumn{2}{c}{3.293 (4.685)}\\\hline
&$\gamma$&\multicolumn{2}{c}{1.620 (1.123)}&&\multicolumn{2}{c}{1.520 (1.054)}&&\multicolumn{2}{c}{1.860 (1.030)}\\
$\sigma^2=0.8^2$&$l_2(\bm \beta)$&\multicolumn{2}{c}{2.697 (2.564)}&&\multicolumn{2}{c}{3.263 (3.226)}&&\multicolumn{2}{c}{4.894 (6.106)}\\\hline
\end{tabular}
\end{table}

\begin{figure}
\centering
\begin{subfigure}[b]{0.65\textwidth}
\includegraphics[width=\linewidth]{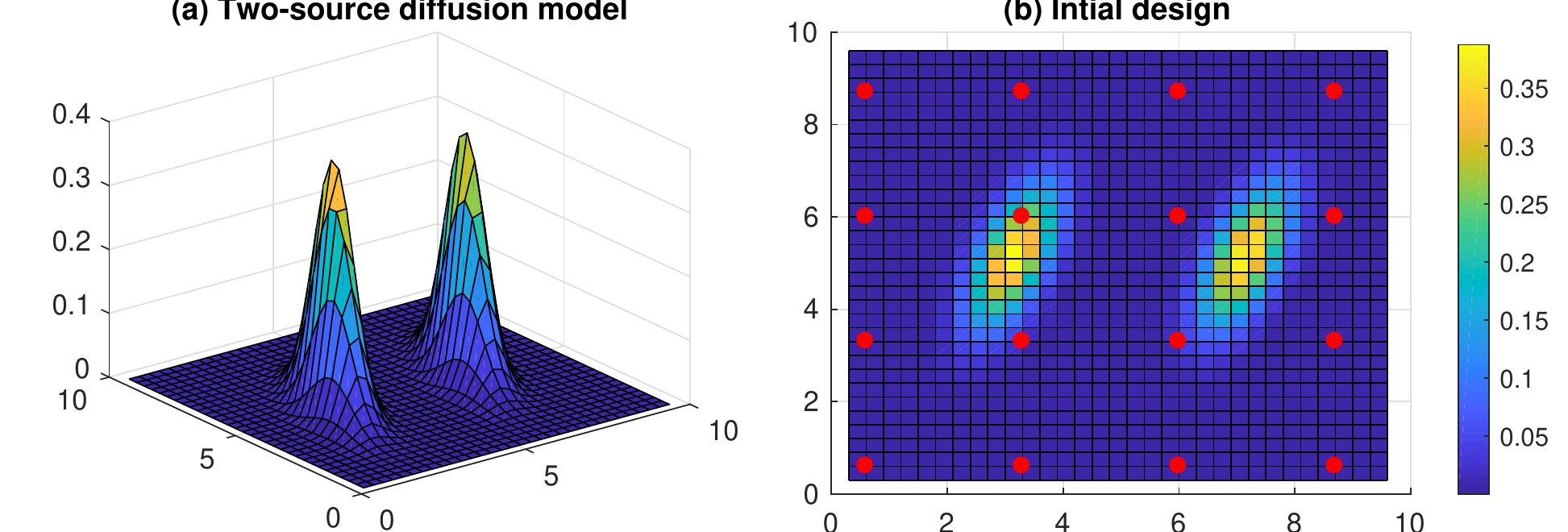}
\caption*{(I) Reference}
\end{subfigure}
~
\begin{subfigure}[b]{0.65\textwidth}
\includegraphics[width=\linewidth]{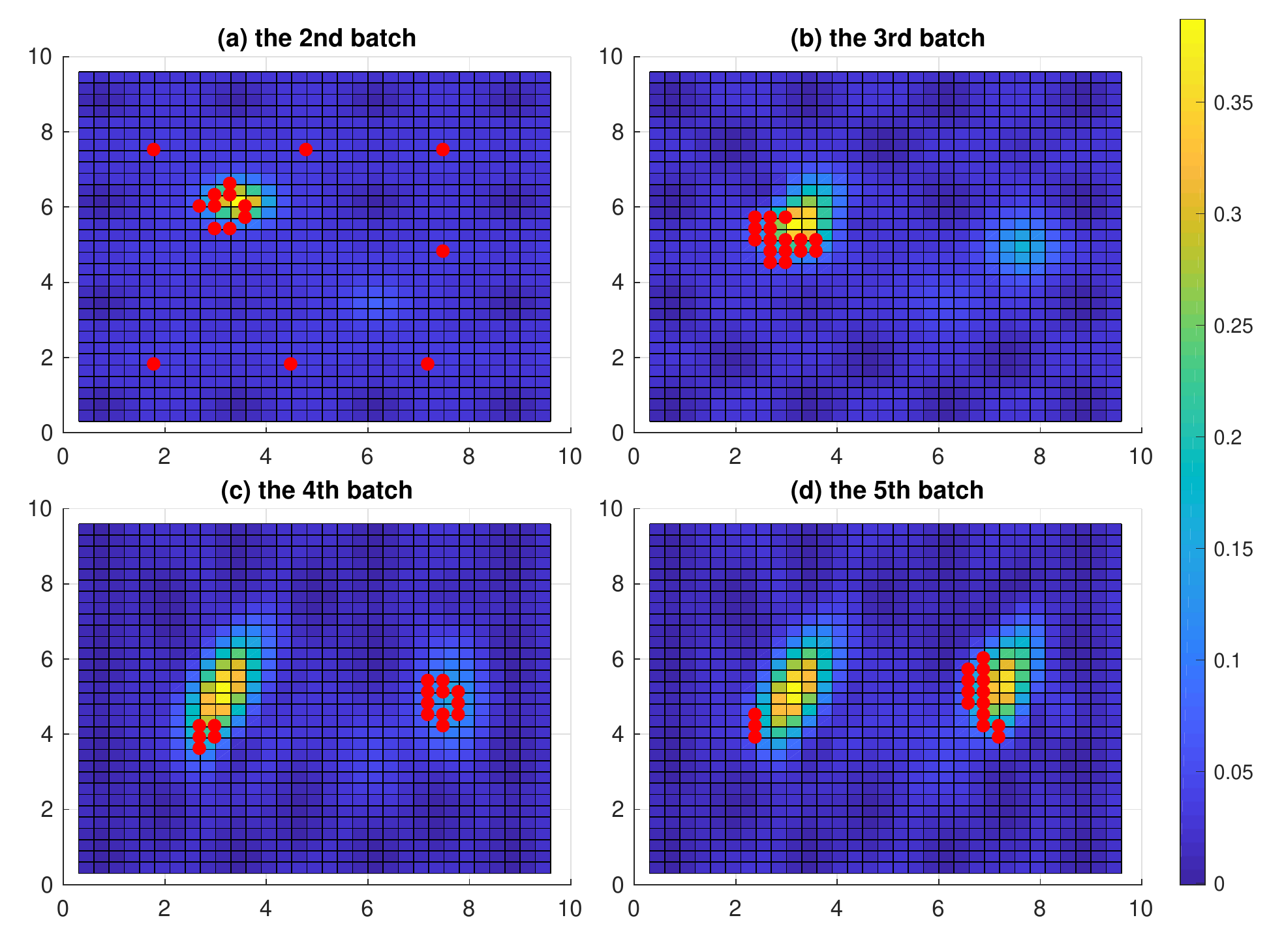}
\caption*{(II) ACDS}
\end{subfigure}
~
\begin{subfigure}[b]{0.65\textwidth}
\includegraphics[width=\linewidth]{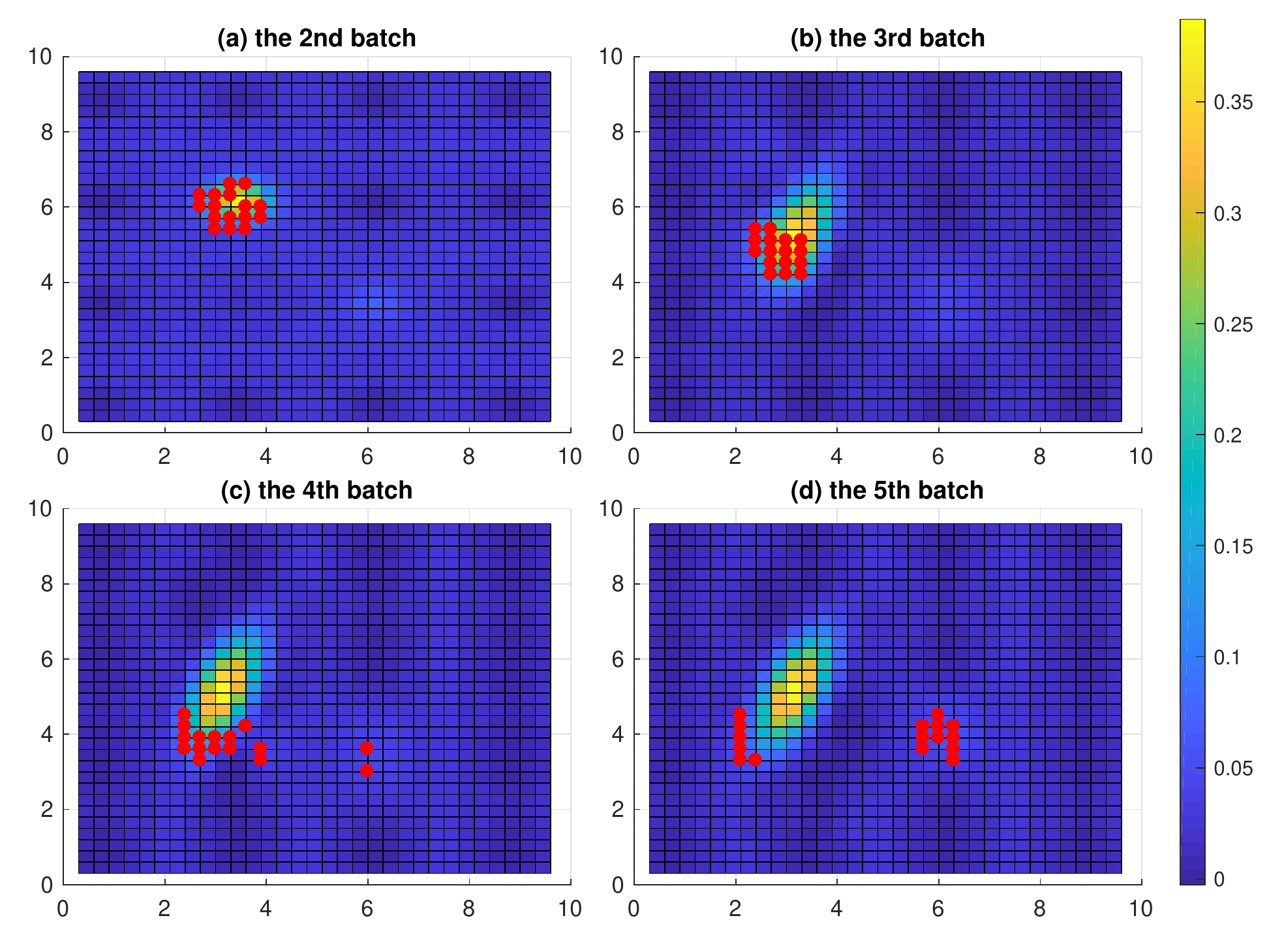}
\caption*{(III) D-optimality}
\end{subfigure}
\caption{Air pollution: four snapshots of sequentially added batch of design points for $\sigma^2=0.2^2$ by using ACDS and D-optimal.}
\label{fig:heat2}
\end{figure}

Table \ref{tab:heat3} compares ACDS, D-optimal and maximin space-filling designs.
We simulate over 50 random trials and compare the mean and standard deviation of $\gamma$ and $l_2(\bm \beta)$.
Both ACDS and D-optimal design approaches are much more accurate than the maximin space-filling design.
Although D-optimal design achieves almost similar accuracy compared with ACDS, it sometimes fails to identify the position of two sources as shown in Figure \ref{fig:heat2} (III).
We also compare ACDS with D-optimal design with a particular initial design generated by the mesh grid shown in Figure \ref{fig:heat2} (I)-(b).
As shown in Figure \ref{fig:heat2} (II) and (III), if the initial design has missed covering one of two hills, ACDS distributes the design points around the missing hill in the next several sequential, whereas D-optimal would concentrate around just one hill.
Therefore D-optimal is more sensitive to the initial design.
From the above comparison, the ACDS outperforms both D-optimal and space-filling designs.

\begin{figure}
\centering
\includegraphics[width=\linewidth]{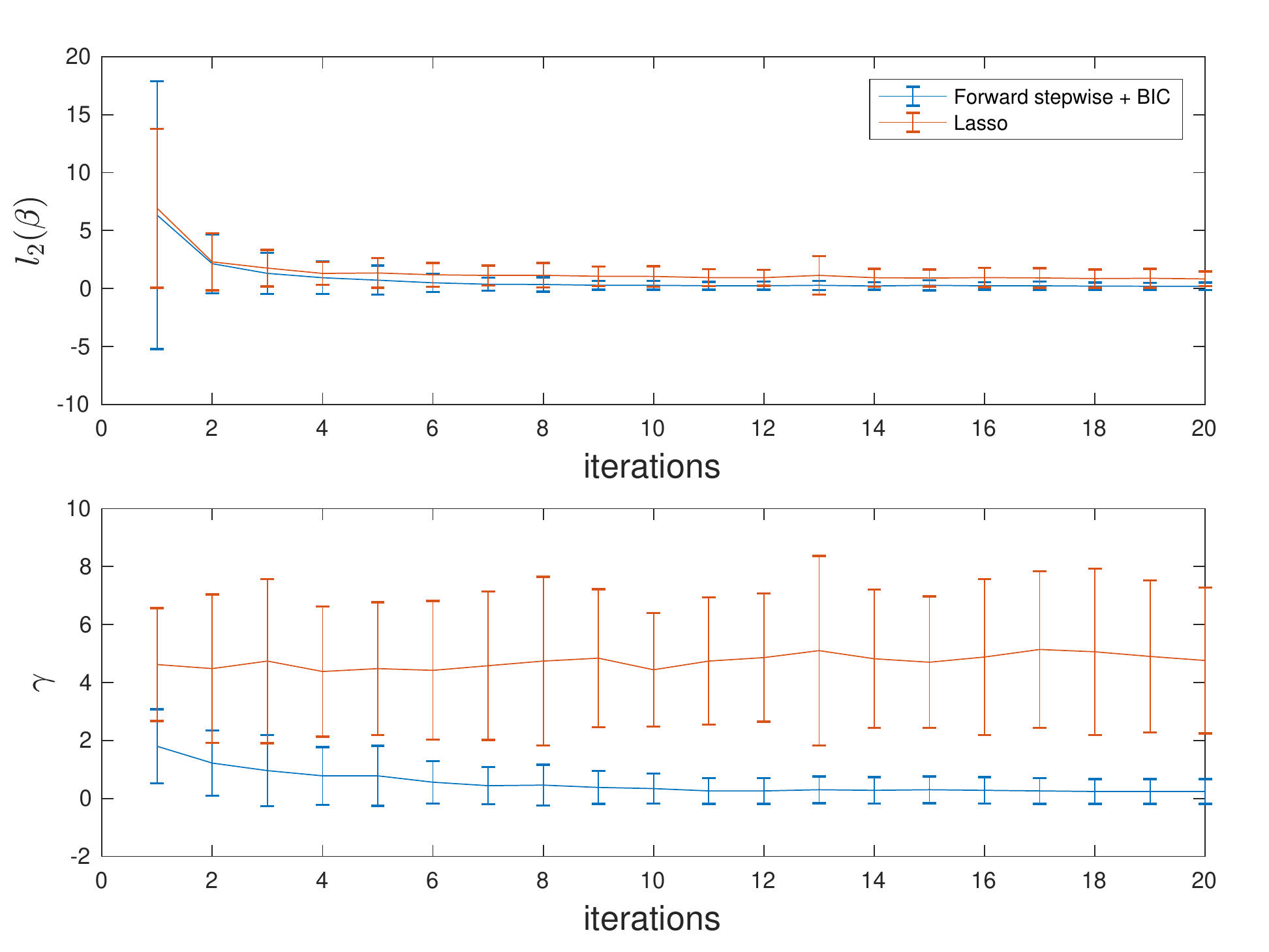}
\caption{Air pollution: the comparison between the forward stepwise regression with BIC and the Lasso approach. The two measures $l_2(\bm \beta)$ and $\gamma$ are changed with respect to the iteration of the active learning algorithm. The blue solid line connects the means of $l_2(\bm \beta)$ and $\gamma$ values of 50 simulations returned by the forward stepwise regression with BIC method and the red solid line represents the Lasso method. The two ends of each vertical line segment are one standard deviation above and below the mean.}
\label{Lasso}
\end{figure}

Lastly, we compare different variable selection methods based on motivating examples.
We consider forward stepwise regression with BIC criteria and Lasso, which are implemented in MATLAB functions \verb|stepwisefit| and \verb|lasso|.
Figure \ref{Lasso} are boxplots of $l_2(\bm \beta)$ and $\gamma$ over 50 random trials.
In terms of $l_2(\bm \beta)$ the two are equally good, but the forward stepwise regression plus BIC has a much smaller number of misidentified terms, measured by $\gamma$.

\section{Remarks on Case Studies}

To summarize the numerical studies in both Section \ref{sec:simu} and \ref{sec:air}, we observe the following advantages of the proposed active learning procedure with the ACDS criterion.
\begin{enumerate}
\item Accuracy. If terminated when the convergence is reached, the proposed method is more likely to identify the correct terms of the differential equations with parameters closer to the truth, compared with the space-filling design.
\item Data economy. Although in Table \ref{tab:ODE1} the space-filling design uses less data on average than the proposed method, it is not as accurate as of the proposed method.
In fact, from our experience in running these simulations, to achieve the same level of accuracy in terms of $l_2(\bm \beta)$ and $\gamma$, the sample size of space-filling design and D-optimal design must be significantly larger, and the algorithm has to use smaller $Tol$ or terminates at a fixed large sample size $N$.
\item Variable selection method. Compared with the existing literature method, as in \cite{zhang2018robust} and \cite{schaeffer2017learning} and others mentioned referred in Section 1, the variable selection we used is much simpler.
We believe that the ACDS criterion and the sequential learning both have made the variable selection method more accurate.
\end{enumerate}

In these case studies, numerical solvers have to be used first to solve the equations, which use fine grids in time $t$ and space $\bm x$ to apply the finite-difference scheme to obtain the $u$, $u_t$, $u_x$,..., and then we add the noise to construct the simulated data.
Thus, the selected design points in $t$ and $\bm x$ have to be from the fine grid-points and cannot be as flexible as in the common physical experiments.
In fact, we have applied the proposed active learning method to select a subset of the complete outputs of the numerical solvers.
However, it is important to point out that the proposed method is being demonstrated as a sequential design method in the case studies, rather than sequential sampling, because we do not use the observed data of the potential design points when deciding which new batch of points are to be selected.
Therefore, the proposed active learning can be used in a real physical experiment, in which the data are truly collected sequentially.
Unfortunately, we do not have a real case study for illustration at this moment.

For a PDE system, the active learning method selects design points in $\bm x\in \mathbb{R}^p$, and it does not matter whether the design points are selected in increasing order in some dimensions.
For an ODE system, the sequential design is in terms of the variable $t$.
If $t$ denotes something other than time, such as one-dimensional location, then the proposed method can be used to learn the ODE system from a physical experiment.
However, if the variable $t$ means time in the physics sense, the sequentially added design points must be increasing in value because time only travels in one direction.
So in each iteration, the active learning needs to select the time points of the future, and the surrogate model must be an accurate forecasting model.
The stationary GP model cannot be applied here.
Users must consider other proper stochastic time series model as the surrogate, which is a question we would investigate in the future.
Alternatively, using the proposed method, the experimenter needs to repeat the experiment to collect a new batch of data at the selected time points.

\section{Discussions}

In this work, we propose an active learning approach with adaptive design criteria combining the D-optimality and maximin space-filling criterion to learn the unknown differential equations from the noisy experimental data.
The Gaussian process model is used as the surrogate model to replace the unknown function when the ACDS criterion is computed for the potential design points.
The weights combining the D-optimality and the space-filling criterion are data-driven, and the active learning procedure is completely autonomous.
Through three simulation case studies, we show the proposed approach is better than the space-filling design and the sequential D-optimal design in terms of two different performance measures on the accuracy of the estimated differential equations.

The proposed method cannot be used to learn the initial and boundary conditions of the partial differential equations, as we only use the observations at $t=t_s$ and the observations on the boundary are not necessarily available.
Hence, we do not need the GP surrogate model to meet the unknown boundary conditions.
On the other hand, if the boundary conditions are known to the experimenter, the GP models should be fitted with the boundary conditions as shown in \cite{tan2018gaussian}.
Consequently, the GP models would be closer to true solutions.

There are several possible avenues for future work.
As pointed out in Section 6, to learn the time-dependent ODE system from physical experiments, we need to find another stochastic model of time series that can produce accurate forecasting.
Some other combinations of design criteria can be combined, such as A/I-optimality for the regression model and other space-filling or prediction based criteria for the GP model.


\bibliographystyle{asa}
\bibliography{ref}
\end{document}